\documentclass{COMNET}

\usepackage[sort&compress]{natbib}
\bibpunct{[}{]}{,}{n}{,}{;}
\usepackage{amsmath}
\usepackage{graphicx} 
\usepackage{algorithmic}
\usepackage{pdflscape}
\usepackage{xcolor}
\usepackage{afterpage}
\usepackage{times,amsmath}
\usepackage{latexsym,amssymb,lscape,multirow}
\usepackage{bm}
\usepackage{array}
\usepackage{url}
\usepackage{multirow}

\begin{document}

\title{Comparing directed networks via denoising graphlet distributions}


\author{
\name{Miguel E. P. Silva$^*$}
\address{Department of Computer Science, University of Manchester, UK\email{$^*$Corresponding author: msilva@dcc.fc.up.pt}}
\name{Robert E. Gaunt}
\address{Department of Mathematics, University of Manchester, UK}
\name{Luis Ospina-Forero}
\address{The Alliance Manchester Business School, University of Manchester, UK}
\name{Caroline Jay}
\address{Department of Computer Science, University of Manchester, UK}

\name{Thomas House}
\address{Department of Mathematics, University of Manchester, UK}}

\maketitle

\begin{abstract}
{Network comparison is a widely-used tool for analyzing complex systems, with
applications in varied domains including comparison of protein interactions or
highlighting changes in structure of trade networks. In recent years, a number
of network comparison methodologies based on the distribution of graphlets
(small connected network subgraphs) have been introduced. In particular, NetEmd
has recently achieved state of the art performance in undirected networks. In
this work, we propose an extension of NetEmd to directed networks and deal with
the significant increase in complexity of graphlet structure in the directed
case by denoising through linear projections. Simulation results show that our
framework is able to improve on the performance of a simple translation of the undirected NetEmd algorithm to the directed case, especially when networks
differ in size and density.}
{directed networks; network comparison; network topology; principal component
analysis; independent component analysis}
\end{abstract}

\section{Introduction}

Complex networks represent relationships between actors of systems with non-trivial structural properties and are ubiquitous in a myriad of domains. Studying the networks underlying these complex systems is then a step towards gaining an understanding about the systems themselves. Comparing different objects is a fundamental part of human cognition, therefore it is only natural that analysis of networks encompasses a comparison element. It is sometimes straightforward to show two networks are different, through finding a mismatch between the set of nodes and edges that compose the networks, which would  and claim a difference if these sets do not match. On the other hand, knowing if two networks are exactly the same is known as the graph isomorphism problem, which has been shown to be in the NP complexity class~\cite{cook1971complexity}. Between these two extremes, \emph{network comparison} has been established as an area of study within network analysis that combines network statistics and computable properties of a network to tell how similar or dissimilar two networks are. Network comparison has been the subject of an increasing amount of research, see for example~\cite{wegner2017identifying, sarajlic2016graphlet, aparicio2016extending, ali2014alignment,tantardini2019comparing}, and its applications are widespread, most notably in biological areas like protein-protein interaction networks~\cite{ali2014alignment} and metabolic networks~\cite{aparicio2016extending}, but also in other domains, like tracking dynamics of world trade networks~\cite{yaverouglu2014revealing}.

Network comparison methods can be broadly split in two categories: node alignment methods~\cite{kuchaiev2010topological,mamano2017sana,gu2018homogeneous}, that attempt to create a mapping between the nodes of the networks being compared, and alignment-free methods~\cite{milo2004superfamilies,prvzulj2007biological,shervashidze2011weisfeiler,onnela2012taxonomies,koutra2013deltacon, ahmed2017graphlet}, that use global features of the networks to determine similarity or dissimilarity. Among the latter, methods that use distributions of small connected subgraphs, known as \emph{graphlets}, have emerged as the state of the art in the area~\cite{prvzulj2007biological,yaverouglu2014revealing,ali2014alignment,aparicio2016extending,sarajlic2016graphlet}. In particular, Wegner et al.~\cite{wegner2017identifying} recently introduced the NetEmd measure that achieves state of the art performance in undirected networks, by comparing the shape of graphlet distributions using the \emph{Earth Mover's Distance} (EMD).

Undirected networks represent relationships between entities under the assumption that the relationship is symmetrical. For example, a Facebook friendship is mutual because both users mutually agree to become friends and thus there is no difference between source and target in the friendship. This abstraction is often insufficient to capture more nuanced relationships. For instance, in Twitter there is an asymmetry in the follower-followed relationship because the action of following another user is not necessarily reciprocated. Directed networks allow us to model such relationships, but this increased wealth of information leads to greater complexity when analyzing them. Such is the case when comparing directed networks, where network comparison metrics designed for undirected networks have been shown to be unsuitable for the task~\cite{sarajlic2016graphlet,aparicio2016extending}.

The first contribution of our work is an extension of NetEmd to directed networks, using directed graphlets of size up to 4. Xu and Reinert~\cite{xu2018triad} previously proposed a similar extension, named \emph{TriadEMD}, that is limited to graphlets of size 3. The main obstacle when going from size 3 to size 4 graphlets in directed networks is the combinatorial explosion in number of orbits due to edge direction, there are 33 orbits in graphlets of size up to 3 but 730 when including size 4 as well. In the undirected case, scaling up from size 3 to size 4 is a comparatively much smaller jump, from 4 to 15. We propose two methodologies for our extension to size 4: the first is a simple extension, where all orbits are used in the NetEmd calculation; in the second, we adapt the idea of Apar\'icio et al.~\cite{aparicio2016extending} and use only orbits that have non-zero frequency in both networks for the comparison. We test systematically the performance difference between using size 3 and size 4 graphlets to understand under which circumstances the added computational load of size 4 graphlets leads to gains in comparison performance. It is also worth noting that the previous methods that use graphlets for comparing directed networks~\cite{sarajlic2016graphlet,aparicio2016extending} also use size 4 graphlets as input to their comparison measures.

Another difference between our work and \emph{TriadEMD} is the inclusion of orbits from size 2 graphlets in the comparison, which is tantamount to including a comparison of the degree distribution in the network comparison measure. Degree distributions are the most commonly studied property to measure the structure of networks and are able to distinguish between networks created with different models~\cite{ravasz2003hierarchical,newman2006structure,prvzulj2004modeling}, but often they are insufficient alone as a heuristic for network comparison as it is possible to craft networks with the same degree distribution but widely different structures~\cite{prvzulj2004modeling,li2005towards}. The inclusion of orbits from size 2 graphlets is also supported by the undirected version of NetEmd, where these orbits are also included~\cite{wegner2017identifying}.

Complex networks are heterogeneous in the amount of data they represent, with famous examples ranging from Zachary's karate club~\cite{zachary1977information} with only 34 individuals to gigantic networks like the Friendster social network~\cite{yang2015defining}, which contains more than one billion relationships between its users. Such disparity makes the task of comparing networks even harder, as networks that differ in size by multiple orders of magnitude may still be organized according to similar topography, for instance when they are generated by the same process. Identifying such ``common organizational principles"~\cite{wegner2017identifying} is the basis for the NetEmd comparison measure. However, networks representing real world phenomena are often inaccurate representations of that phenomena due to unseen data, measurement errors or simply because the systems they represent are too complex to fully describe. These inaccuracies lead to noise in the statistics used to describe the network and similar topographies become harder to recognize, especially when networks differ greatly in size.

The distributions of graphlets are not immune to these errors, so our second contribution in this work is using denoising methods to reduce the impact of these errors so that NetEmd is able to more accurately distinguish networks according to those core structures that compose each network. To this end, we propose a framework that employs Principal Component Analysis (PCA)~\cite{hotelling1933analysis} as the most commonly used denoising method before comparing the graphlet distributions with EMD. PCA finds uncorrelated directions of ordered by contribution to variance, which is sufficient for independence under the assumption of normality~\cite{shlens2014tutorial}. Graphlet distributions for complex networks, however, break this assumption as, although hard to characterize exactly, they are thought to be similar to degree distributions, which, whether or not they are approximately power-law as often claimed~\cite{barabasi1999emergence}, are at least often far from normal due to being heavy-tailed~\cite{broido_scale-free_2019}. Therefore, we hypothesize that PCA struggles to create an appropriate model for these distributions, which can make the denoising process ineffective, so we also propose an alternative intermediate step that uses Independent Component Analysis (ICA)~\cite{comon1994independent}, a method that is explicitly designed to work with heavy-tailed distributions \cite{mackay2003information}.

We test our contributions in clustering tasks involving synthetic and real world networks, following the same experimental procedure as the original NetEmd paper~\cite{wegner2017identifying}, which adheres to the framework of graph comparison introduced by Yavero{\u{g}}lu et al.~\cite{yaverouglu2015proper}. We find that our proposed method of denoising using component analysis techniques is able to improve NetEmd's performance on clustering tasks, both in undirected and directed networks. This improvement is particularly significant when comparing networks of different sizes and densities, where NetEmd is already the state of the art.

\section{Background}

\subsection{Graph theoretical concepts}

A \emph{graph} $G$ is composed of a set of \emph{vertices} (or \emph{nodes}) $V(G)$ and a set of edges $E(G) \subseteq V(G)\times V(G)$, represented by pairs $(a,b) \in E(G)$ for $a, b \in V(G)$. A graph can be either $directed$, when the order of the vertices in the pairs expresses direction, meaning that $(a,b)\in E(G)$ does not imply that $(b,a)\in E(G)$, or \textit{undirected} otherwise (i.e.\ when $(a,b)\in E(G)$ if and only if $(b,a)\in E(G)$). The \emph{size} of a graph is the number of vertices in the graph, written as $|V(G)|$, and the \emph{density} is the portion of edges present in the graph over the total potential ones ($|E(G)|/\binom{|V(G)|}{2}$ in undirected networks). For a directed network, the \textit{reciprocity} $\rho$ is the probability that, given an edge $(a,b)$ selected uniformly at random from $E(G)$, $(b,a)$ is also a member of $E(G)$. A graph is called \emph{simple} if it does not contain multiple edges (two or more edges connecting the same pair of vertices) or self-loops (an edge of the form $(a,a)$ that connects a vertex to itself). In this work, we consider only simple graphs.

The \emph{neighbourhood} of a vertex $u \in V(G)$ is defined as $N(u) = \{ v : (v,u) \in E(G) \lor (u,v) \in E(G) \}$. All nodes are assigned consecutive integer numbers starting from $0$ and running to $|V(G)|-1$. The \emph{degree} of a vertex is the number of edges it participates in, which is equivalent to the size of the node's neighbourhood. In directed networks, the degree of a vertex can be split into the \emph{in-degree} and \emph{out-degree}, counting the number of incoming and outgoing edges, respectively, from that node. In directed networks, the \emph{degree} of a vertex is the sum of its \emph{in-degree} and \emph{out-degree}. In undirected networks, the \emph{degree} of a node $u$ is simply $|N(u)|$. The distribution of degrees in a network is called the \emph{degree sequence}.

A \emph{subgraph} of size $k$, $G_k$, of a graph $G$ is a graph with $k$ vertices, such that $V(G_k) \subseteq V(G)$ and $E(G) \supseteq E(G_k) \subseteq V(G_k) \times V(G_k)$. A subgraph is \emph{induced} if $\forall u,v \in V(G_k), \; (u,v) \in E(G_k) \leftrightarrow (u,v) \in E(G)$ and is said to be \emph{connected} when all pairs of vertices have a sequence of edges connecting them. Two graphs $G$ and $H$ are \emph{isomorphic}, written as $G\sim H$, if there is a bijection between $V(G)$ and $V(H)$ such that two vertices are adjacent in $G$ if and only if their correspondent vertices in $H$ are adjacent. A \emph{match} of a graph $H$ in a larger graph $G$ is a set of nodes that induce the respective subgraph $H$. In other words, it is a subgraph $G_k$ of $G$ that is isomorphic to $H$. The frequency of a subgraph $G_k$ is then the number of different matches of $G_k$ in $G$.

\emph{Orbits} are unique positions of a graph, calculated by partitioning the set of vertices into equivalence classes where two vertices belong to the same class if there is an automorphism that maps one into the other~\cite{ribeiro2019survey}.

Small, connected, non-isomorphic, induced subgraphs are commonly called \emph{graphlets}~\cite{prvzulj2007biological}. The smallest graphlet considered is a single edge, which can be seen as a size-2 subgraph. An undirected edge has a single orbit (three in the directed case) and the frequency of a node in this orbit is equivalent to the degree of the node. The \emph{graphlet degree vector} of a node is an extension of the definition of degree to size-$k$ graphs, representing how many times the node occurs in each orbit. The \emph{graphlet degree matrix} of a graph is the collection of graphlet degree vectors of each node in the graph.

\begin{figure}
    \includegraphics[width=0.95\textwidth]{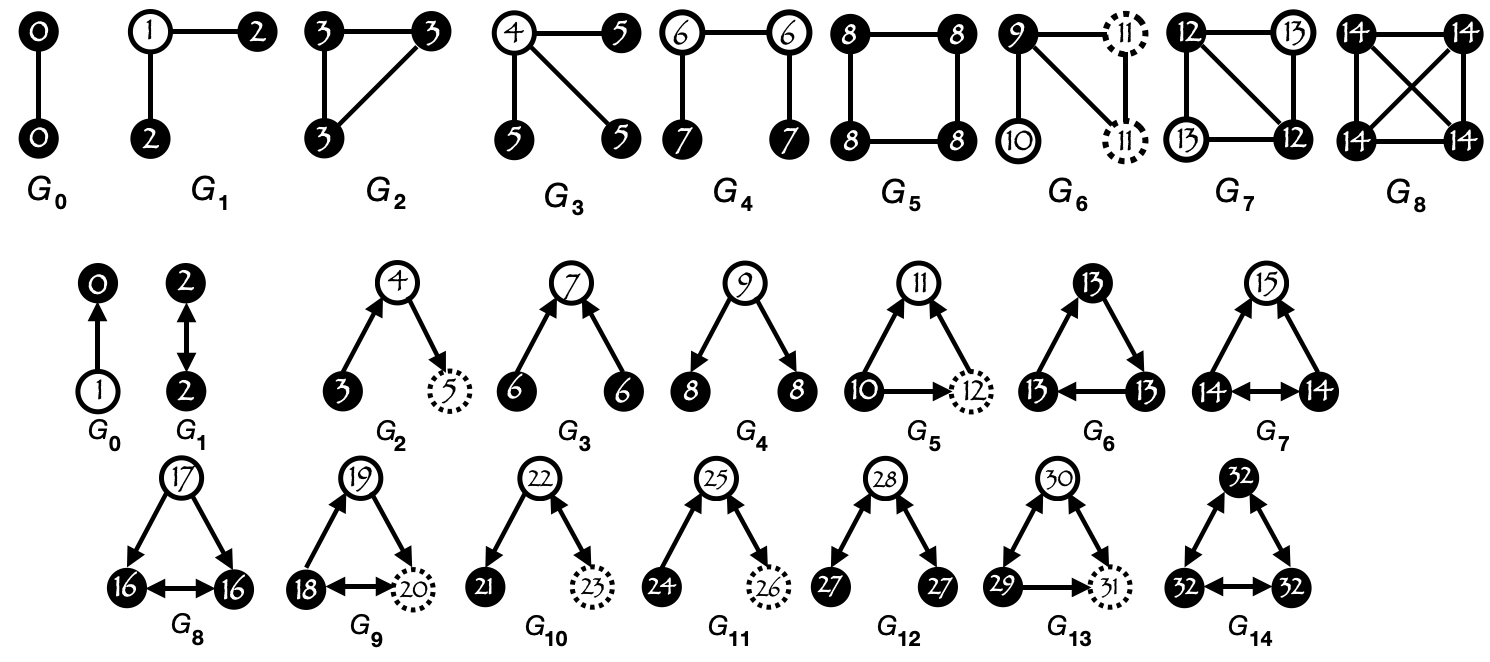}
    \caption{Undirected graphlets of size 2, 3 and 4 and directed graphlets of size 2 and 3. Nodes are numbered according to their orbit, counted by type (directed or undirected) from the size 2 graphlets, and nodes with the same number represent the same orbit.}
    \label{fig:graphlets}
\end{figure}

Figure~\ref{fig:graphlets} shows the graphlets of size 2, 3 and 4 in undirected networks and size 2 and 3 in directed networks, alongside the respective orbits.

\subsection{NetEmd}

NetEmd~\cite{wegner2017identifying} is a network comparison measure that relies on structural features of the network, mainly the distribution of orbit frequencies. The core idea is formalizing the intuition that the shape of the degree distribution is indicative of the network's generation mechanisms, for instance, a network with a power law degree distribution is generated by a process distinct from a network with a uniform degree distribution. As the graphlet degree vector is a generalization of the degree distribution for graphlets of size $k \geq 3$, the shapes of the distributions of each orbit also carry information about the topology of the network. Note that because the shape of a distribution is invariant under linear transformations such as translations, using the shape as the focus of the comparison is well-suited to comparing networks of different sizes and densities. 

Wegner et al.~\cite{wegner2017identifying} postulate that ``any metric that aims to capture the similarity of shapes should be invariant under linear transformations of its inputs." Thus, they define a measure of similarity between distributions $p$ and $q$, with non-zero and finite variances, using the Earth Mover's Distance (EMD)~\cite{rubner1998metric}:

\begin{equation*}\label{eq:emd}
EMD^{*}(p,q) = \text{inf}_{c \in \mathbb{R}} (EMD(\tilde{p}(\cdot + c), \tilde{q}(\cdot))),
\end{equation*}
where $\tilde{p}$ and $\tilde{q}$ are the distributions resulting of scaling $p$ and $q$ to variance 1. Any distance metric $d$ can be used to generate $d^*$; EMD was used by the authors as it has been shown to be an appropriate metric to compare shapes of distribution in domains such as information retrieval and it produced better results than other distance metrics, like the Kolmogorov or $L^1$ distances.

Given two networks $G$ and $H$ and a set of $m$ orbits $\mathcal{O} = \{o_1, o_2, \ldots, o_m\}$, the \emph{NetEmd} measure is defined as:

\begin{equation}
\label{eq:netemd}
NetEmd_{\mathcal{O}}(G,H) = \frac{1}{m} \sum\limits_{i=1}^{m} EMD^*(p_{o_i}(G),p_{o_i}(H)),
\end{equation}
where $p_{o_i}(G)$ and $p_{o_i}(H)$ are the distributions of orbit $i$ in graphs $G$ and $H$, respectively. Note that $\mathcal{O}$ may be replaced by any set of network features; in this work we focus on orbits only.

\section{Directed NetEmd}

Directed networks pose challenges unlike the ones observed in undirected networks, due to the combinatorial explosion in number of orbits introduced by distinguishing $(u,v)$ and $(v,u)$ as different edges between $u$ and $v$. There are 730 orbits when considering directed graphlets of size up 4, compared to 15 in the undirected case, and scaling up to size 5 in directed networks becomes unfeasible as the number of orbits rises to $45,637$. This sharp increase makes the task of counting orbit frequencies for each node even harder, with no known analytical approaches like ORCA~\cite{hovcevar2014combinatorial} that rely on crafting sets of equations that exploit combinatorial relationships between smaller graphlets to compute orbit counts. Instead, enumeration-based approaches are required in the directed case, which are known to be at least an order of magnitude slower than analytical approaches in undirected networks. In order to adapt NetEmd to directed networks, we use the G-Trie~\cite{ribeiro2014g} data structure and the counting algorithm proposed by Apar\'icio et al.~\cite{aparicio2016extending}, publicly available at~\cite{gtscanner}, modified to return the graphlet degree matrix instead of the graphlet degree distribution.

\begin{figure}[t]
    \centering
    \includegraphics[width=0.45\textwidth]{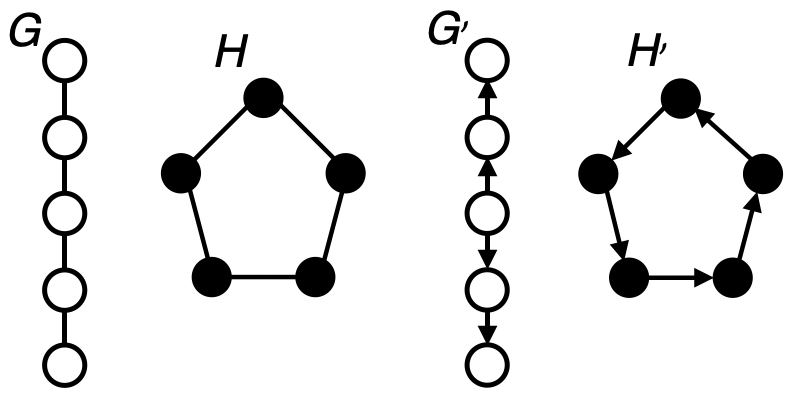}
    \caption{Example to illustrate the usefulness of only using orbits that appear in at least one of the networks being compared. Graphs $G$ and $H$ are the example provided by Apar\'icio et al.~\cite{aparicio2016extending}; we adapt them to directed networks $G'$ and $H'$, keeping the same number of orbits in each network. We measure $NetEmd(G',H') = 5e^{-5}$ and $Weighted\_NetEmd(G',H') = 0.12$.}
    \label{fig:weight_by_orbit_example}
\end{figure}

How to handle this increased number of orbits in the comparison is also a challenging problem, especially when comparing networks that contain only a small subset of orbits and differences within those orbits can get diluted when taking the average over the whole set. Apar\'icio et al.~\cite{aparicio2016extending} argue that the comparison should only be done over the orbits present in at least one of the networks, using networks $G$ and $H$ in Figure~\ref{fig:weight_by_orbit_example} as an example. In this case, the authors show that using the original formulation of the Graphlet Degree Distribution Agreement (GDA) (described in Section~\ref{sec:gda}) inflates the similarity score between the two networks far beyond what would be expected from two so distinct looking networks, with an agreement score of 0.92 against 0.32 when using their modified version that compares only orbits in at least one of the networks. This modification to GDA is well-founded because GDA is a measure of \emph{agreement}, bounded between 0 and 1 and meant to be interpretable; a score close to 1 means that the networks are similar and a score close to 0 the opposite, so the magnitude of the value that the measure outputs is of interest. However, NetEmd is a measure of distance between networks and the relative difference between distances is more informative than the absolute value of that distance, so including information that an orbit is missing in both networks improves our ability to tell if they are less distant from each other when comparing against another pair of networks.

With the above in mind, we propose two versions of directed NetEmd. Given two networks $G$ and $H$ and a set of $m$ orbits $\mathcal{O} = \{o_1, o_2, \ldots, o_m\}$, the first uses all $m$ orbits in this set, which in practice are all the orbits in graphlets of size up to 3 or 4 and the formulation is the same as the original NetEmd in Equation~\ref{eq:netemd}. The second is using the same idea of Apar\'icio et al.~\cite{aparicio2016extending}, which we refer to as \emph{weighted NetEmd}, where we restrict the set of orbits to those that occur in at least one of the networks $G$ or $H$. This leads to a new set of $m'$ orbits $\mathcal{O'} = \{o_1', o_2', \ldots, o_m'\}$, with $m' \leq m$. The average in Equation~\ref{eq:netemd} is done over this set $\mathcal{O'}$ instead of the original $\mathcal{O}$, but the rest of the formula is the same:

\begin{equation}
\label{eq:weighted_netemd}
Weighted\_NetEmd_{\mathcal{O'}}(G,H) = \frac{1}{m'} \sum\limits_{i=1}^{m'} EMD^*(p_{o_i}(G),p_{o_i}(H)).
\end{equation}

Graphs $G'$ and $H'$ in Figure~\ref{fig:weight_by_orbit_example} show an analogous example to the one shown by Apar\'icio et al.~\cite{aparicio2016extending} to demonstrate the difference between weighted and original NetEmd applied to directed networks. Although it is usually impractical to scale up to size 5 in directed networks, in a small example like this one it becomes feasible to do so. In this context, we calculate a distance of $5e^{-5}$ using all orbits and $0.12$ when using the weighted version. Note that this idea of using the orbits that appear in at least one of the networks under comparison can also be applied to undirected networks, as Apar\'icio et al~\cite{aparicio2016extending} do, however we show results only for the directed case.

\section{NetEmd with Dimension Reduction}
\label{sec:dimred}

\subsection{Motivation}

In this section, we describe how we couple dimensionality reduction techniques with NetEmd, using them as noise reduction techniques. The rationale is that processes that generate networks are inherently noisy due to the complexity of the systems they represent. This is particularly evident when we consider random networks generated by the same model; we expect them to have similar structure and that similarity to be reflected in the distribution of subgraph and orbit frequencies, but the stochastic nature of the generation process introduces noise in these distributions that makes them harder to identify.

Considering the graphlets and orbits shown in Figure~\ref{fig:graphlets}, it is clear that (for example) directed graphlet $G_2$ cannot exist if $G_0$ does not, and similarly directed graphlet $G_{12}$ cannot exist if $G_1$ does not. More generally, it is clear that the different orbits do not represent independent degrees of freedom, and are expected to be constrained by the requirement of combinatorial consistency of the underlying full graph $G$ as well as correlations induced by the generation process.

In practice, both combinatorial enumeration of graphlets and \textit{a priori} determination of statistical relationships between them (beyond comparison to simple random graphs as in \cite{milo2004superfamilies}) are computationally infeasible. While this does not matter in practice for the undirected case, for directed graphs as we consider here, the relatively large number and complexity of orbits makes these much more important. We therefore seek an approach that can adjust for the non-independence of orbit counts.

Proceeding somewhat formally, let $\mathbf{F}_G$ denote the graphlet degree matrix of graph $G$ for a set of orbits $\mathcal{O}$, a $n \times m$ matrix where $n = |V(G)|$, $m = |\mathcal{O}|$, such that $[\mathbf{F}_{G}]_{i,j}$ represent the frequency of orbit $j$ for node $i$. Further, let $\mathbf{f}_i$ be the vector of frequencies of node $i$, i.e.\ $(\mathbf{f}_{i})_{j} = [\mathbf{F}_{G}]_{i,j}$ meaning $\mathbf{f}_i$ is the $i$-th row of $\mathbf{F}_G$. Now note that we are only interested into networks up to isomorphism; in fact, our distance measures depend on empirical histograms of orbit counts, deriving the distribution $p_{o_i}$ in \eqref{eq:netemd} from histogram heights like
\begin{equation}
    h_{o_j,y} = \frac{1}{n}\sum_{i=1}^{n} 
    \mathbf{1}_{\{(\mathbf{f}_{i})_{j} = y\}} ,
    \label{hoiy}
\end{equation}
which is the proportion of nodes with $y$ counts of orbit $o_j$, where $\mathbf{1}$ is an indicator function taking the value 1 if its argument is true and the value 0 otherwise. The right-hand side of \eqref{hoiy} is a sum of random variables, one for each node, and from \eqref{eq:netemd} we will further sum over orbits for a function of two such histogram heights.

In general, we will expect a random graph model to assign a probability to each graph in some finite set, and according to this measure we expect a probability distribution to be induced on $\mathbf{h}_{G} = [h_{o_j,y}]$. As discussed above, we expect both combinatorial constraints and correlations between orbit counts, meaning that a sum over all elements of $\mathbf{h}_{G}$ as implied by \eqref{hoiy} and \eqref{eq:netemd} will inflate, due to the additivity of variances of random variables, the amount of noise in EMD realisations compared to the amount that is absolutely necessary under the random graph model.

Therefore, a natural approach is to denoise using techniques that do not require explicit solution of or simulation from random graph models. In particular, we will use the linear techniques of principal component analysis (PCA) and independent component analysis (ICA), noting their computational efficiency compared to potentially more general nonlinear methods \cite{Hinton:2006}. The main idea of our methodology is therefore to project the orbit frequencies to a lower dimension, training the dimension reduction model to lose minimal information while removing noise.

Since it is not guaranteed that each graph will dimensionally reduce to the same size, the approaches we use allow for expansion of the dimensionally reduced features back to the original (high) dimension. The NetEmd comparison from Equation~\ref{eq:netemd} is consequently applied to the reconstructed frequencies. By using the first $L$ components of the linear methods, which contain the most signal from the dataset, to reconstruct the original counts, this can be intuitively understood as decreasing noise within the orbit frequencies when we do not have a good model for that noise.

\subsection{Principal Component Analysis}

Principal component analysis (PCA) is a technique for dimensionality reduction that preserves as much variability of the data as possible, by computing principal components of the data. Principal components are sequences of orthogonal unit vectors that form an orthonormal basis, in which the original dimensions of the data are linearly uncorrelated. When projecting the data onto this new basis, these vectors can be seen as the directions that maximize the variance of the projected data, with the first principal component representing the maximum variance.

Let $\mathbf{F}_G$ denote the graphlet degree matrix of graph $G$ for a set of orbits $\mathcal{O}$, a $n \times m$ matrix where $n = |V(G)|$, $m = |\mathcal{O}|$ and $[\mathbf{F}_{G}]_{i,j}$ represents the frequency of orbit $j$ for node $i$. We assume that the frequencies have been normalized, a preprocessing step advised before applying PCA (to prevent notation overload, we use the same $\mathbf{F}_G$ to denote the normalized version of the graphlet degree matrix). The principal components of this matrix are defined by $\mathbf{V} = \mathbf{F}_G\mathbf{W}$, where $\mathbf{W}$ is a $m \times m$ matrix whose columns are the eigenvectors of $\mathbf{F}_G^\intercal\mathbf{F}_G$. To allow for only the first L components, the matrix $\mathbf{W}$ can be truncated to $\mathbf{W}_L$, with dimensions $m \times L$, leading to a transformation $\mathbf{V}_L = \mathbf{F}_G\mathbf{W}_L$. This truncation is done such that the $L$ eigenvectors chosen to correspond to the largest eigenvalues of $\mathbf{W}$. The original graphlet degree matrix can be reconstructed as $\hat{\mathbf{F}}_G = \mathbf{V}_L\mathbf{W}_L^\intercal = \mathbf{F}_G\mathbf{W}_L\mathbf{W}_L^\intercal$. The goal of PCA is to learn $\mathbf{W}_L$ such that the variance of the original data preserved is maximized, while also minimizing the total squared reconstruction error $||\mathbf{F}_G - \hat{\mathbf{F}}_G||_2^2 = ||\mathbf{V}\mathbf{W}^\intercal - \mathbf{V}_L\mathbf{W}_L^\intercal||_2^2$.

We use the reconstructed graphlet degree matrix to compare networks with NetEmd, keeping Equation~\ref{eq:netemd} virtually unchanged from the original formulation:

\begin{equation}
\label{eq:pca_netemd}
PCA\_NetEmd_{\mathcal{O}}(G,H) = \frac{1}{m} \sum\limits_{i=1}^{m} EMD^*(\hat{p}_{o_i}(G),\hat{p}_{o_i}(H)),
\end{equation}
where $\hat{p}_i(G)$ and $\hat{p}_i(H)$ are obtained from $\hat{\mathbf{F}}_G$ and $\hat{\mathbf{F}}_H$ respectively, instead of
$\mathbf{F}_G$ and $\mathbf{F}_H$.

\subsubsection{Choosing the number of components.}
\label{subsec:num_comps}

Choosing an appropriate number of components to project the data down affects the reconstruction error and therefore the amount of noise reduced. If we take $L = m$, then clearly we are able to reconstruct the original frequencies perfectly and no noise has been removed from the data. On the other hand, picking a value for L too low, leads to foregoing descriptiveness of the data. A common strategy for picking the number of components is through the amount of variance explained by each component, a strategy that allows us to adapt the number of components to suit the networks we are comparing.
  
The sample covariance matrix of each orbit, i.e. $F_{G_{i,1}}, \, F_{G_{i,2}}, \, \ldots, \, F_{G_{i,m}}$, is proportional to $\mathbf{F}_G^\intercal\mathbf{F}_G$ (the derivation for this result can be found in~\cite{jolliffe2002springer}, pp.~30-31) and contains the variance of the frequency of each orbit in its diagonal. As a square matrix, according to the spectral theorem, the covariance matrix can be diagonalized by its eigenvectors and the values of the resulting diagonal matrix are the corresponding eigenvalues. These eigenvalues represent variability of each axis in the projected space. Therefore, by taking the sum of the $L$ highest eigenvalues, the ones corresponding to the first $L$ components, we get the variance explained by the first $L$ principal components. The ratio $\sum_1^L \lambda_i / \sum_1^m \lambda_i$, where $\lambda_i$ is the eigenvalue corresponding to the $i$th eigenvector, measures the proportion of variance explained by the first $L$ principal components. Finally, to condition the number of components on the proportion of variance explained, the smallest $L$ is calculated such that $\sum_1^L \lambda_i / \sum_1^m \lambda_i \geq r$ where $r \times 100\%$ is the percentage of variance explained.

\subsection{Independent Component Analysis}

Independent component analysis (ICA) is a statistical technique in which an observed vector of random variables is thought to be the linear combination of unknown independent components. These components are assumed to be mutually statistically independent and with zero mean. The classic example of an ICA application is a party where a microphone is picking up multiple conversations and the goal is to separate the source signal from each conversation from the mixed one picked up by the microphone.

Transposing ICA to the domain of network comparison, we can draw an analogy from the above application by considering that the voices in the conversation are the nodes in the network and the orbit frequencies are the data points captured by the microphone. Our goal is then to search for the source signals, i.e., network characteristics stemming from its generation mechanism that generate such frequency distribution. We conjecture that by using these source signals to reduce noise in the orbit frequency distributions, our network comparison measure becomes able to more accurately distinguish networks with different generation mechanisms.

As before, let $\mathbf{F}_G$ denote the graphlet degree matrix of graph $G$ for a set of orbits $\mathcal{O}$ and let $\mathbf{f}_i$ be the vector of frequencies of node $i$, i.e. $f_{i,j} = \mathbf{F}_{G_{i,j}}$. We can write $\mathbf{f}_i$ as $\mathbf{f}_i = \mathbf{A}\mathbf{s}_i$, where $\mathbf{A}$ is called the mixing matrix and $\mathbf{s}_i$ are the $L$ independent components. The goal of ICA is to estimate $\mathbf{A}$ and $\mathbf{s}_i$ from $\mathbf{f}_i$ only. In practice, algorithms to calculate independent components compute a weight matrix $\mathbf{W}$, a pseudo-inverse of $\mathbf{A}$ and obtain the independent components through $\mathbf{s}_i = \mathbf{W}\mathbf{f}_i$. Using the FastICA algorithm~\cite{hyvarinen2000independent,hyvarinen1999fixed}, $\mathbf{W}$ is constructed iteratively by finding unit vectors $\mathbf{w}$ such that the projection $\mathbf{w}^\intercal\mathbf{f}_i$ maximizes non-gaussianity, measured by the approximation of negentropy $J(\mathbf{w}^\intercal\mathbf{f}_i) \propto \left[ E(G(\mathbf{w}^\intercal\mathbf{f}_i)) - E(G(\mathbf{\nu})) \right]^2$, where $G(x) = \log\cosh x$ and $\mathbf{\nu}$ is a standard Gaussian random variable with mean 0 and variance 1. These unit vectors $\mathbf{w}$ are combined into the weight matrix $\mathbf{W}$ and decorrelated to prevent convergence to the same maxima by $\mathbf{W} = \mathbf{W} / \sqrt{\| \mathbf{W}\mathbf{W}^\intercal \|}$ (the matrix norm is calculated using the $l^2$ norm); this procedure is then repeated with $\mathbf{W} = \frac{3}{2}\mathbf{W} - \frac{1}{2} \mathbf{W} \mathbf{W}^\intercal\mathbf{W}$ until convergence.

Upon obtaining the weight matrix, we calculate its pseudo-inverse to obtain the mixing matrix $\mathbf{A}$ and calculate $\hat{\mathbf{F}}_G$ similarly to what was done with PCA: we project $\mathbf{F}_G$ to a lower dimensional space using $\mathbf{W}$ and apply the mixing matrix $\mathbf{A}$ to retrieve the original data. The final NetEmd formula then takes a similar form as for $PCA\_NetEmd$:

\begin{equation}
\label{eq:ica_netemd}
ICA\_NetEmd_{\mathcal{O}}(G,H) = \frac{1}{m} \sum\limits_{i=1}^{m} EMD^*(\hat{p}_{o_i}(G),\hat{p}_{o_i}(H)),
\end{equation}
with $\hat{p}_i(G)$ and $\hat{p}_i(H)$ obtained from $\hat{\mathbf{F}}_G$ and $\hat{\mathbf{F}}_H$ respectively, as before.

\begin{figure}
    \centering
    \includegraphics[width=\textwidth]{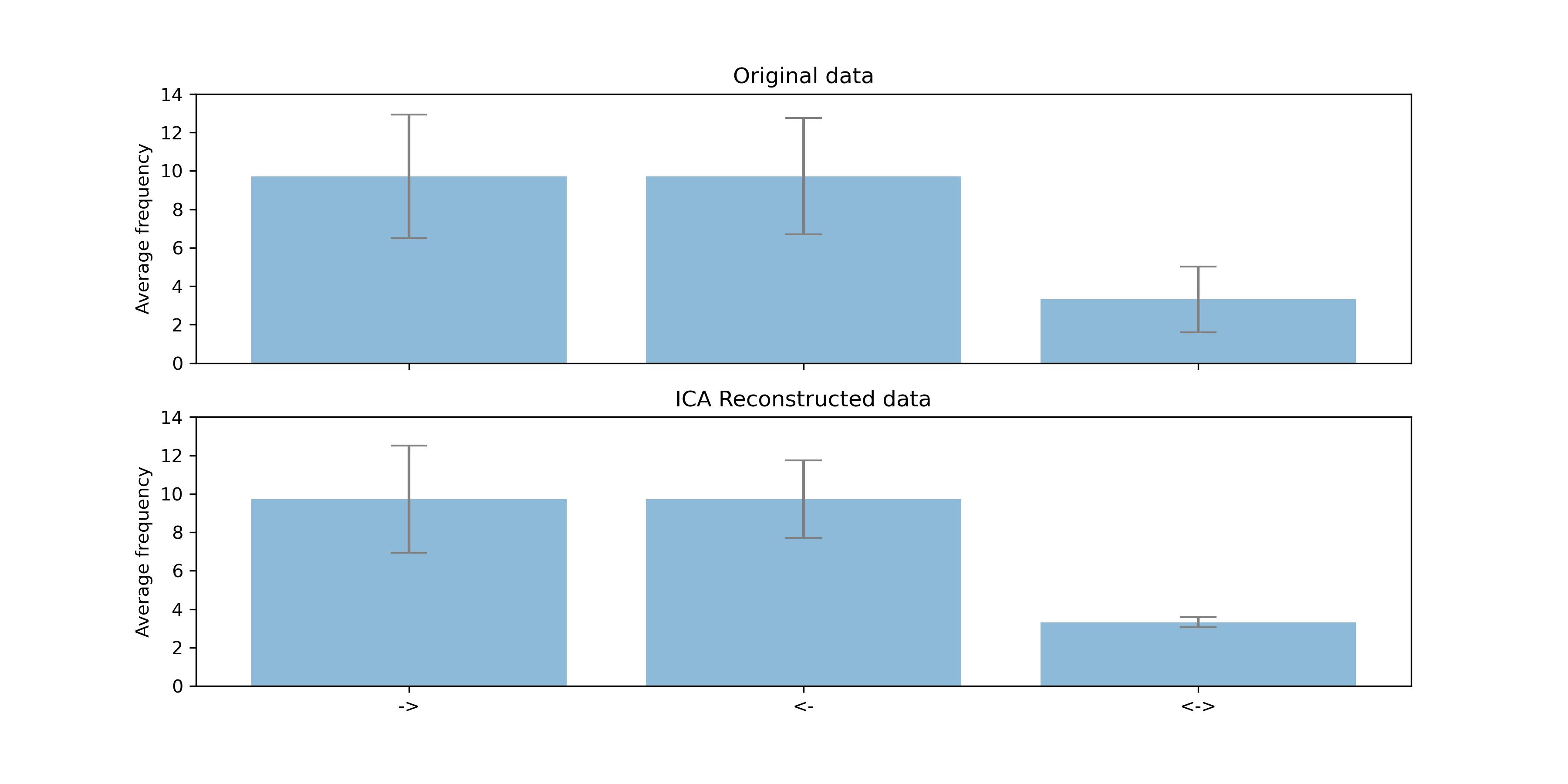}
    \caption{Example of the impact of ICA denoising in the distributions of size 2 directed graphlets, in a network with 50 nodes, generated via a directed ER model, with $p = 0.25$.}
    \label{fig:ica_example}
\end{figure}

As we mention previously, our hope is that the independent components discovered are more representative of the true number of degrees of freedom involved in generating the network than the full number of orbits. Although a formal argument for this intuition is beyond scope of this work, we present here a straightforward example that can help clarify the impact of this denoising process on the orbit frequency distribution. Consider an Erd\H{o}s-R{\'e}nyi (ER)~\cite{erdHos1960evolution} random network model, adapted for directed networks, where all edges are independent and exist with probability $p$ (thus, $\text{Pr}((u,v) \in E(G)) = p, \forall u, v \in V(G), u \neq v$). We can derive analytically the expected frequency of orbits of size 2 for each node in this network, corresponding to the number of outgoing, incoming and reciprocal edges, which are $p(1-p)(n-1)$, $p(1-p)(n-1)$ and $p^2(n-1)$, respectively. When generating networks from this model, we empirically observe stochastic noise generating values away from these frequencies, particularly when the number of nodes in the network is small and these frequencies deviate from their asymptotic behaviour. For this simple model, we denoise via a single independent component, with results shown in Figure~\ref{fig:ica_example}. This shows the impact of noise reduction in a network with 50 nodes, created using the model above, and in particular very significant reduction in the noise associated with reciprocal edges.

\section{Experimental Setup}

Experiments were performed on an AMD Opteron Processor 6380 with 1.4 GHz and 2 MB of cache memory, using Python version 3.5.4. The implementation of our methods is available online at \url{https://github.com/migueleps/denoise-dir-netemd}.

\subsection{Measure of Cluster Performance}

Given a set of networks $\mathcal{G} = \{G_1, G_2, \ldots, G_N\}$, divided in disjoint classes $C = \{c_1, c_2, \ldots, c_m\}$, we use the performance measure proposed by Wegner et al.~\cite{wegner2017identifying} that captures the idea that networks of the same class should be nearer to each other than to networks of other classes. It is defined as an empirical probability $P(G) = d(G,G_1) < d(G,G_2)$, where $G_1$ is a network selected randomly from the same class as $G$ and $G_2$ is randomly selected from a different class, and $d$ is the network comparison statistic. The performance over the whole dataset is the average $P(G)$ over all the networks in $\mathcal{G}$, and we write this as $\overline{P} = P(\mathcal{G}) = \frac{1}{|\mathcal{G}|} \sum _{G \in \mathcal{G}} P(G)$. In Appendix~\ref{app:diff_metrics}, we present results using area under precision-recall curve (AUPR) and adjusted Rand index (ARI).

\subsection{Synthetic datasets}

We reproduce the experimental setup of Wegner et al.~\cite{wegner2017identifying}, testing our proposed modifications to NetEmd on the same eight random network models. These models are Erd\H{o}s-R{\'e}nyi~\cite{erdHos1960evolution}, Barabasi-Albert preferential attachment~\cite{barabasi1999emergence}, configuration model, geometric random graphs~\cite{gilbert1961random}, geometric gene duplication model~\cite{higham2008fitting}, duplication divergence of Vazquez et al.~\cite{vazquez2003modeling}, duplication divergence of Ispolatov et al.~\cite{ispolatov2005duplication} and Watts-Strogatz~\cite{watts1998collective}. Details of parameters for each model are described in Appendix~\ref{app:network_models}.

We generate 10 networks per model per combination of number of nodes and average degree. The set of number of nodes used is $N \in \{1250, 2500, 5000, 10000\}$ and the set of average degrees is $k \in \{10, 20, 40, 80\}$. This leads to a total of 160 networks per model and 1280 networks in total. 

For the directed datasets, we add varying levels of reciprocity $\rho \in \{0, 0.25, 0.5, 0.75, 1\}$. To generate a directed network with reciprocity $\rho$, we take the undirected version, we duplicate and invert all edges (if $(u,v)$ is in the undirected network, we add $(v,u)$ to the directed version) and we then take a proportion $1-\rho$ of these pairs and choose randomly a direction to remove (either $(u,v)$ or $(v,u)$). This is done incrementally such that if $(u,v)$ is removed from the dataset with 75\% reciprocity, then it is also removed from the datasets with 50\%, 25\% and 0\% reciprocity.

These datasets of synthetic networks give rise to two tasks aimed at gauging how well a network comparison measure separates clusters according to their generation mechanism, with the ground truth given by the random network model used to generate the networks in each cluster. The first, more simple task is to separate networks with the same number of nodes and the same density. To this end, we create 16 groups based on the combinations of $N$ and $k$, each group containing 10 realizations of each random network model, for a total of 80 networks. This task is equivalent to $RG_1$ in the original NetEmd paper~\cite{wegner2017identifying}. The second task, equivalent to $RG_3$ in the original NetEmd paper, is to compare all 1280 networks simultaneously, finding the 8 clusters of 160 networks. This task measures how sensitive the comparison measure is to differences in an order of magnitude in number of nodes and edges, making the separation by model type more difficult. In directed networks, we repeat Task 1 and 2 for each level of reciprocity, determining how sensitive the network comparison is to this third parameter that impacts the set of orbits available for comparison (the set of orbits with 0\% reciprocity is disjoint from the set of orbits with 100\% reciprocity).

\vspace{-5pt}
\subsection{Real world datasets}

We use the dataset of Onnela et al.~\cite{onnela2012taxonomies} to validate our methodology on a mix of real world and synthetic networks. The multiple sources that make up this dataset lead to a heterogeneous set of networks; however, the ability to separate these networks according to their domain is desired from a network comparison method. We were unable to find sources for the original 746 networks, so we use the reduced set proposed by Ali et al.~\cite{ali2014alignment} with 151 unweighted and undirected networks. There is no ground truth for the true clusters in this dataset and the dendograms constructed through the methods proposed by Onnela et al.~\cite{onnela2012taxonomies} and Ali et al.~\cite{ali2014alignment} disagree on the composition of each cluster. Therefore, we aim to reconstruct clusters according to the type of data, which can be visualized in the supplementary material of Ali et al.~\cite{ali2014alignment}. 

For directed networks, we download four datasets from the SNAP library~\cite{snapnets}: Gnutella peer-to-peer file sharing network from August 2002~\cite{ripeanu2002mapping,leskovec2007graph} (9 networks with an average reciprocity of 0\%), CAIDA autonomous systems relationships datasets from January 2004 to November 2007~\cite{leskovec2005graphs} (122 networks with an average reciprocity of 100\%), ego networks of \emph{circles} from Google+~\cite{mcauley2012learning} (132 networks with an average reciprocity of 28\%) and ego networks of \emph{lists} from Twitter~\cite{mcauley2012learning} (973 networks with an average reciprocity of 54\%). 

\subsection{Other network comparison methods}

\subsubsection{GCD.}

Yavero{\u{g}}lu et al.~\cite{yaverouglu2014revealing} note that orbit counts have dependencies between them, making them redundant since they can be expressed as a linear combination of other orbits. The authors identify 11 out of 15 orbits in graphlets of size up to 4 and 56 out of 73 orbits in graphlets of size up to 5 as non-redundant. The authors construct the graphlet degree vector of each node in the network using this reduced number of orbits (although the full set of orbits may also be used), making up a matrix where each row is the graphlet degree vector of each node, and compute the Spearman's correlation coefficient between each pair of orbits, i.e., each pair of columns in this matrix. The pairwise Spearman's correlation coefficients are aggregated in a square $m \times m$ matrix called the \emph{Graphlet Correlation Matrix} (GCM), where $m$ is the number of orbits used (e.g., 11 if using non-redundant orbits of graphlets up to size 4).

To compare two networks, the authors propose the \emph{Graphlet Correlation Degree} (GCD), which is the Euclidean distance between the upper triangle of the GCM of each network.

Sarajli{\'c} et al.~\cite{sarajlic2016graphlet} extend GCD to directed networks, finding 13 non-redundant orbits in graphlets of size up to 3 and 129 in graphlets of size up to 4.

\subsubsection{GDA.}
\label{sec:gda}

Przulj et al.~\cite{prvzulj2007biological} introduced the Graphlet Degree Distribution (GDD), calculated as follows. Let $d^o_G(k)$ be the number of nodes of graph $G$ that participate $k$ times in orbit $o$, i.e., $d^o_G(k)$ is graphlet degree distribution of orbit $o$. The authors scale the distribution to decrease the contributions of larger orbits by calculating $S^o_G(k) = d^o_G(k)/k$ and then normalize the distribution as $N_G^o(k) = S^o_G(k) / \sum_{k=1}^{\infty}S_G^o(k)$. To compare the GDD distributions of the same orbit in different networks $G$ and $H$, the authors propose the GDD-agreement (GDA) metric defined as:

\begin{equation*}
    \label{eq:gda}
        A^o(G,H) = 1 - \left(\sum\limits_{k=1}^{\infty} \left[ N_G^o(k) - N_H^o(k)\right]^2\right)^{1/2}.
\end{equation*}

As a measure of \emph{agreement}, the output of GDA is 1 if the distributions are identical, as opposed to NetEmd which is a measure of \emph{distance} meaning it will output 0 for identical distributions. To aggregate the agreements of the multiple orbits, the authors propose either a arithmetic or a geometric mean. We compare against the implementation of Aparicio et al.~\cite{aparicio2016extending}, which uses the arithmetic mean and is available at~\cite{gtscanner}. This implementation uses the idea that if a graphlet has a frequency of 0 in both networks, then the orbits from that graphlet are excluded from the arithmetic mean, for both in the directed and undirected versions.

\section{Results}

\subsection{Undirected Results}

We present the results for Tasks 1 and 2 in undirected networks and the results for the Onnela et al.~\cite{onnela2012taxonomies} dataset in Table~\ref{tab:undir_scores}. The results for Task 1 show the average and standard error for the 16 values of $\overline{P}$, one for each combination of $N$ and $k$. The results for Task 2 and Onnela et al. dataset show the single value of $\overline{P}$ after comparing the 1280 and 151 networks, respectively.

\begin{table}[t]
\centering
  \scriptsize
  
  \renewcommand{\arraystretch}{1.6}
  \begin{tabular}{ccccc}
    \hline \multirow{2}{*}{Algorithm} & \multirow{2}{*}{Parameter} & \multicolumn{2}{c}{Synthetic} & Real \\
    \cline{3-5}  &  & Task 1 & Task 2 & Onnela et al.\\
    \hline Original NetEmd & G5 & \textit{0.995 $\pm$ 0.003}  & 0.918 & 0.890\\
    \hline \multirow{5}{*}{$PCA\_NetEmd$} & 50\% Variance & 0.98 $\pm$ 0.01 & 0.915 & 0.864 \\
    \cline{2-5} & 80\% Variance & 0.98 $\pm$ 0.01 & 0.918 & \textbf{0.893} \\
   \cline{2-5} & 90\% Variance & \textit{0.995 $\pm$ 0.004} & 0.922 & 0.891 \\
    \cline{2-5} & 95\% Variance & \textit{0.993 $\pm$ 0.004} & 0.919 & 0.892\\
    \cline{2-5} & 99\% Variance & \textit{0.995 $\pm$ 0.004} & 0.921 & 0.891 \\
    \hline \multirow{3}{*}{$ICA\_NetEmd$} & 2 components & 0.985 $\pm$ 0.002 & 0.926 & 0.869 \\
    \cline{2-5} & 10 components & \textit{0.994 $\pm$ 0.004} & \bf{0.930} & 0.887 \\
    \cline{2-5} & 15 components & \bf{0.996 $\pm$ 0.003} & 0.929 & 0.890 \\
    \hline
    \hline \multirow{2}{*}{GCD} & 11 orbits & \textit{0.99 $\pm$ 0.01} & 0.888 & 0.789\\
    \cline{2-5} & 73 orbits & \textit{0.994 $\pm$ 0.008} & 0.885 & 0.819\\
    \hline \multirow{3}{*}{GDA} & G3 & 0.986 $\pm$ 0.003 & 0.868 & 0.876 \\
    \cline{2-5} & G4 & 0.977 $\pm$ 0.007 & 0.886 & 0.855 \\
    \cline{2-5} & G5 & 0.95 $\pm$ 0.01 & - & 0.832 \\
    \hline
  \end{tabular}
  \caption{Results for Task 1, Task 2 and the Onnela et al.~\cite{onnela2012taxonomies} datasets in undirected networks. The metric used for Task 1 is the mean (and standard error of the mean) of the 16 values for $\overline{P}$ in each combination of number of nodes with network density. The metric used for Task 2 and Onnela et al. dataset is the sole value of $\overline{P}$ after comparing the 1280 and 151 networks, respectively. The parameter column indicates: graphlet size used in original NetEmd; percentage of variance explained to determine the number of components in $PCA\_NetEmd$ (using orbits in graphlets of size up to 5); the number of components used in $ICA\_NetEmd$, using orbits in graphlets of size up to 5; the number of orbits used by GCD; graphlet sizes used in GDA. The bolded values are the maximum for each task. Italic values are within one standard error of the maximum performance. Note that results for GDA with graphlet size 5 took longer than a week to return results for Task 2, at which point we stopped the computation.}
  \label{tab:undir_scores}
\end{table}

We find that our proposed modifications to NetEmd achieve an increase in performance, with $ICA\_NetEmd$ outperforming original NetEmd in synthetic datasets and $PCA\_NetEmd$ in the Onnela et al. dataset. The performance gain in Task 1 is not significant as the results are within a standard error of each other, but on Task 2, the more difficult task, the difference is more pronounced. This performance improvement in synthetic datasets is not reflected in the Onnela et al. dataset, where 15 components are necessary to equalize the performance of original NetEmd. The best performance in the Onnela et al. dataset is $PCA\_NetEmd$ with 80\% explained variance, which contrasts with the results in the synthetic dataset, where 80\% explained variance achieves worse performances than the original NetEmd. Using higher values for explained variance in $PCA\_NetEmd$ also improves performance over original NetEmd in Task 2 and Onnela et al. dataset, but not in Task 1.

\subsection{Directed Results}

We show the results for Task 1 in directed networks in Table~\ref{tab:scores_dir_task1} and the results for Task 2 and the dataset of real world directed networks in Table~\ref{tab:scores_dir_task2}. The results for Task 1 show the average and standard error for the 16 values of $\overline{P}$, one for each combination of $N$ and $k$, for each level of reciprocity. The results for Task 2 and real world directed networks dataset show the single value of $\overline{P}$ after comparing the 1280, for each level of reciprocity, and 1232 networks, respectively.

Similarly to the undirected case, we find that the results for each algorithm in Task 1 are similar across different levels of reciprocity, meaning that the various different versions of NetEmd and DGCD are able to distinguish the generation mechanism regardless of how many edges are reciprocated in a directed network. Results show that DGCD with 129 orbits achieves best performance for 0\%, 25\%, 50\% and 75\% and $PCA\_NetEmd$ with 90\% explained variance for 100\% reciprocity. We observe no significant difference between using NetEmd with all orbits, with weighted orbits or coupled with dimensionality reduction techniques. Results with TriadEMD are non-significantly better than our NetEmd version with orbits from size 3 graphlets in this task, indicating that using graphlets of size 2 for the comparison in this task does not help differentiating between different models. This is likely due to the configuration model, which shares the same degree distribution as the duplication divergence model of Vazquez et al.~\cite{vazquez2003modeling}.

\begin{table}[t]
\centering
  \scriptsize
  \renewcommand{\arraystretch}{1.6}
  \begin{tabular}{>{\centering\arraybackslash}m{29pt}>{\centering\arraybackslash}m{43pt}>{\centering\arraybackslash} m{16pt}>{\centering\arraybackslash} m{16pt}>{\centering\arraybackslash} m{16pt}>{\centering\arraybackslash} m{16pt}>{\centering\arraybackslash} m{16pt}>{\centering\arraybackslash} m{28pt} }
    \hline \multirow{3}{*}{Algorithm} & \multirow{3}{*}{Parameter} & \multicolumn{5}{c}{Synthetic} & \multirow{3}{*}{\shortstack{Real\\Networks}}\\
    \cline{3-7} & & \multicolumn{5}{c}{Reciprocity} & \\
     &  & 0\% & 25\% & 50\% & 75\% & 100\% & \\
    \hline  \multirow{2}{*}{\shortstack{All\\Orbits}} & G3D & 0.923 & 0.905 & 0.907 & 0.905 & 0.915 & 0.864 \\
    \cline{2-8} & G4D & 0.908 & 0.879 & 0.881 & 0.878 & \textbf{0.927} & 0.813 \\
    \hline \multirow{2}{*}{Weighted} & G3D & 0.918 & 0.899 & 0.902 & 0.898 & 0.913 & \textbf{0.868} \\
    \cline{2-8} & G4D & 0.902 & 0.876 & 0.878 & 0.875 & 0.923 & 0.827 \\
    \hline \multirow{5}{*}{PCA} & 50\% Variance & 0.924 & 0.897 & 0.900 & 0.894 & 0.908 & 0.821 \\
    \cline{2-8} & 80\% Variance & 0.924 & 0.885 & 0.892 & 0.881 & 0.920 & 0.825 \\
    \cline{2-8} & 90\% Variance & 0.921 & 0.882 & 0.889 & 0.878 & \textbf{0.927} & 0.824 \\
    \cline{2-8} & 95\% Variance & 0.915 & 0.880 & 0.887 & 0.876 & 0.922 & 0.822 \\
    \cline{2-8} & 99\% Variance & 0.914 & 0.878 & 0.885 & 0.875 & 0.924 & 0.818 \\
    \hline \multirow{3}{*}{ICA} & 2 components & \textbf{0.941} & \textbf{0.916} & \textbf{0.914} & \textbf{0.911} & \textbf{0.927} & 0.712 \\
    \cline{2-8} & 10 components & 0.934 & 0.900 & 0.903 & 0.896 & 0.923 & 0.843 \\
    \cline{2-8} & 15 components & 0.931 & 0.896 & 0.903 & 0.892 & 0.917 & 0.844 \\
    \hline \multicolumn{2}{c}{TriadEMD} & 0.926 & 0.903 & 0.906 & 0.906 & 0.916 & 0.859 \\
    \hline \multirow{2}{*}{DGCD} & 13 orbits & 0.905 & 0.904 & 0.895 & 0.885 & 0.871 & 0.812 \\
    \cline{2-8} & 129 orbits & 0.887 & 0.884 & 0.880 & 0.880 & 0.885 & 0.799 \\
    \hline \multirow{2}{*}{GDA} & G3D & 0.842 & 0.802 & 0.802 & 0.800 & 0.858 & 0.736 \\
    \cline{2-8} & G4D & - & - & - & - & 0.886 & - \\
    \hline
  \end{tabular}
  \caption{Results for Task 2 in directed networks, for each level of reciprocity, and for the real world networks dataset. The metric used for this task is the sole value of $\overline{P}$ after comparing the 1280 and 1231 networks, respectively. The parameter column indicates: graphlet size used in directed and weighted NetEmd; percentage of variance explained to determine the number of components in $PCA\_NetEmd$ (using orbits in graphlets of size up to 4); the number of components used in $ICA\_NetEmd$, using orbits in graphlets of size up to 4; the number of orbits used by DGCD; graphlet sizes used in GDA. The bolded values are the maximum for each dataset. Note that results for GDA with graphlet size 4 took longer than a week to return results for Task 2 and in the real world networks dataset, at which point we stopped the computation.}
  \label{tab:scores_dir_task2}
\end{table}

In Task 2, our results show a noticeable gain in performance when using $ICA\_NetEmd$ over other NetEmd versions and DGCD, in particular when using only 2 components and in reciprocity levels smaller than 100\%. We find that using orbits in graphlets of only size 2 and 3 orbits achieves better results for this task than using ones with size up to 4, with the exception of the two extremities in reciprocity (0\% and 100\%), where having more available orbits for comparison helps performance. In this task, we also find that using $Weighted\_NetEmd$ degrades performance compared to using all orbits. Results also show that our versions of NetEmd perform very similarly to TriadEMD, with a slight advantage to TriadEMD at 0, 75\% and 100\% reciprocity, which is perhaps expected again due to the configuration model adding a confounding factor when using graphlets $G_0$ and $G_1$ from Figure~\ref{fig:graphlets} in the comparison. 

On the other hand, $Weighted\_NetEmd$ reaches the best performance in the real directed networks dataset, where, similarly to Task 2 in synthetic networks, using smaller graphlets yields better results. Table~\ref{tab:scores_dir_task2} shows that $ICA\_NetEmd$ with 10 and 15 components obtains better performance than NetEmd without dimensionality reduction and with size 4 graphlets. This comparison is unfair to $ICA\_NetEmd$, as results suggest that when there is a large discrepancy between the sizes of the networks, using smaller graphlet sizes improves the performance of NetEmd. Therefore, although $Weighted\_NetEmd$ achieves the best performance out of the parameters we include in Table~\ref{tab:scores_dir_task2}, the top performance in the real world directed networks dataset is achieved by $ICA\_NetEmd$ with size 3 graphlets and 10 components. All four versions of NetEmd that we propose achieve better results than DGCD and GDA in this dataset. This is also true when comparing our versions with size 3 graphlets against TriadEMD, indicating that in real world networks the degree distribution is able to contribute in a positive manner towards being able to discriminate networks of different sources

\vspace{-5pt}
\subsection{Performance by number of ICA components}

We examine the impact on performance of varying the number of components used for $ICA\_NetEmd$, with the caveat that increasing the number of components leads to a greater computational complexity as the FastICA algorithm takes more iterations to converge. We also explore the performance of $ICA\_NetEmd$ under smaller graphlet sizes, namely graphlets sizes 3 and 4 for undirected networks and size 3 for directed. There are two different ways of using smaller graphlets on $ICA\_NetEmd$, the more natural way in which we compute the independent components using the set of orbits related to the desired graphlet size, or an alternative way in which the components are computed using a larger graphlet size and the reconstructed graphlet degree matrix is posteriorly truncated to include only the orbits of the desired graphlet size. For example, when calculating network distances using undirected graphlets of size up to 4, there are 15 orbits. In the first method, the graphlet degree matrix $\mathbf{F}_G$ is a $|V(G)| \times 15$ matrix and we simply calculate $\hat{\mathbf{F}}_G$ as described in Section~\ref{sec:dimred}, constraining the number of components used to a maximum of 14 components. In the second method, $\mathbf{F}_G$ is a $|V(G)| \times 73$ matrix, allowing a maximum of 72 components to compute $\hat{\mathbf{F}}_G$, which is then truncated to 15 columns.

\begin{figure}
   \centering
        \includegraphics[width=0.80\textwidth]{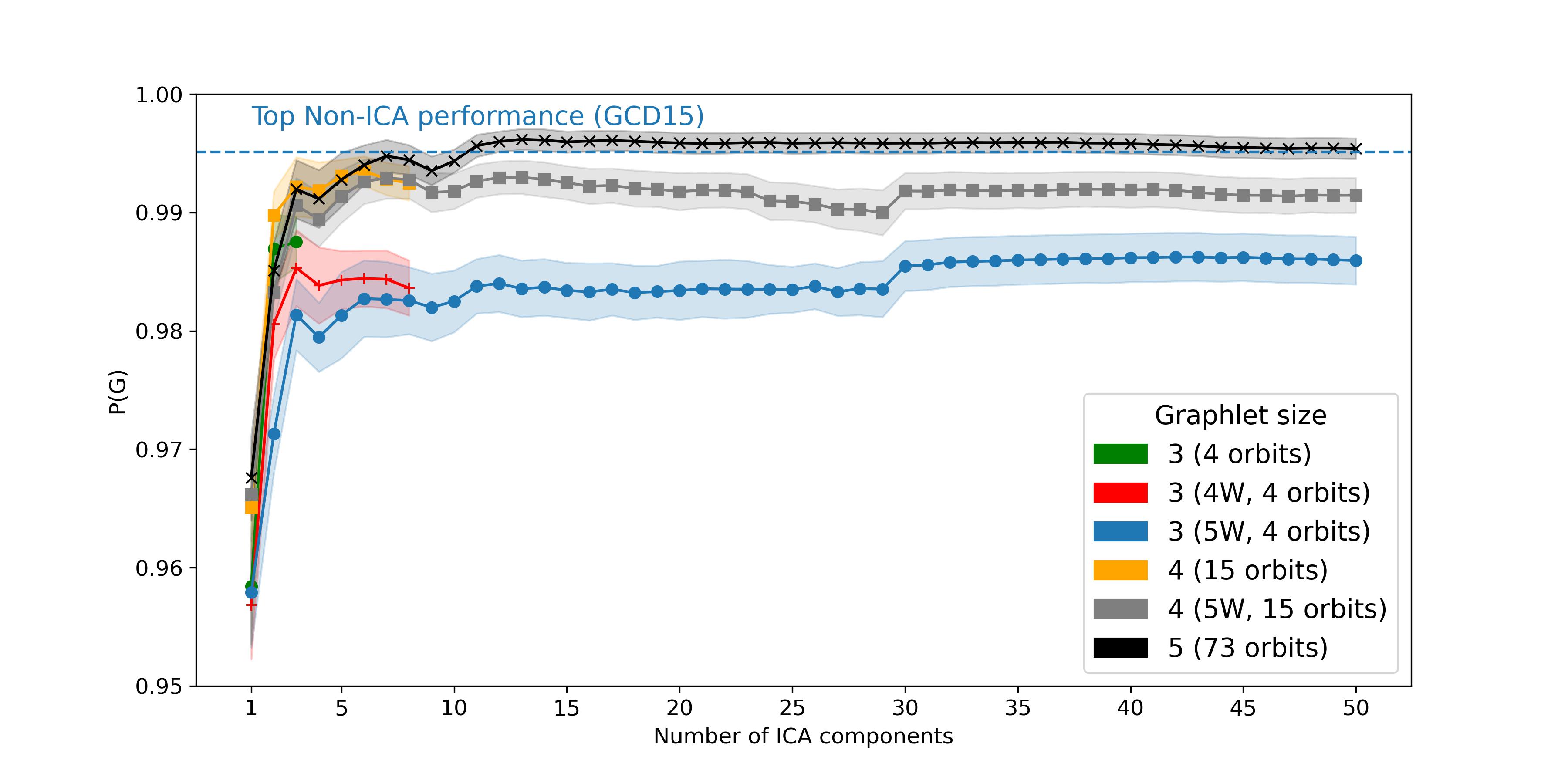}
        \caption{Undirected networks - Task 1}
        \label{fig:undir_task1_pbar}

        \includegraphics[width=0.80\textwidth]{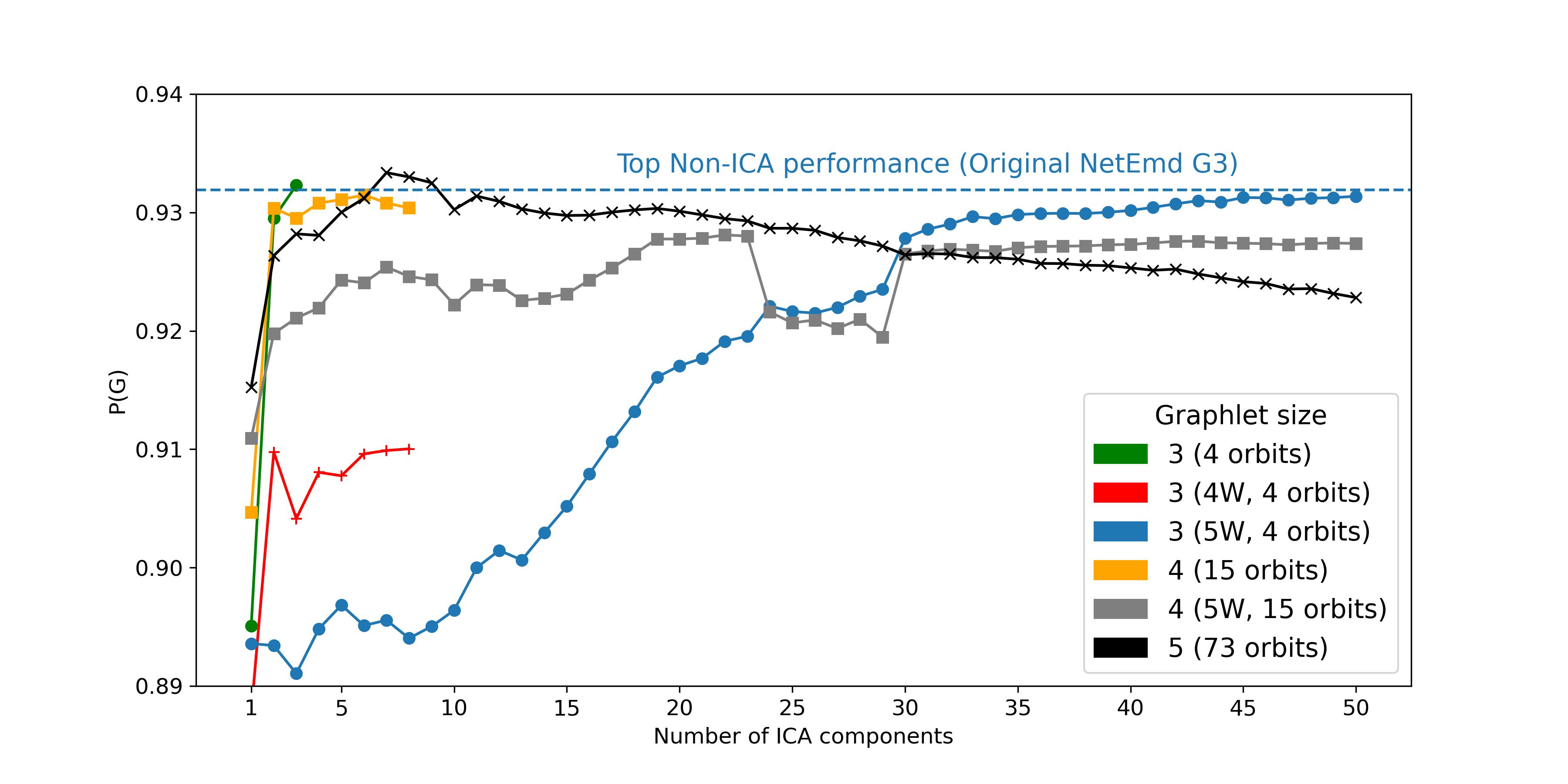}
        \caption{Undirected networks - Task 2}
        \label{fig:undir_task2_pbar}
    
        \centering
        \includegraphics[width=0.95\textwidth]{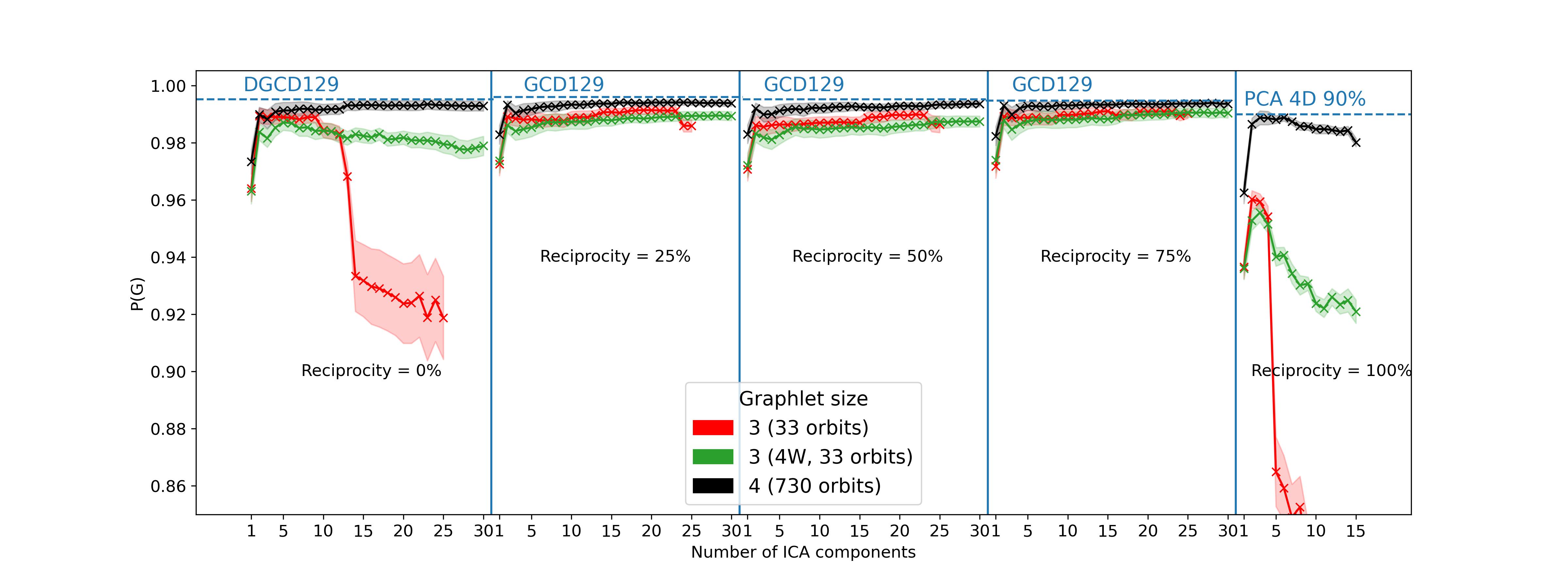}
        \caption{Directed networks - Task 1}
        \label{fig:dir_task1_pbar}
 \end{figure}

\begin{figure}
        \centering
        \includegraphics[width=0.83\textwidth]{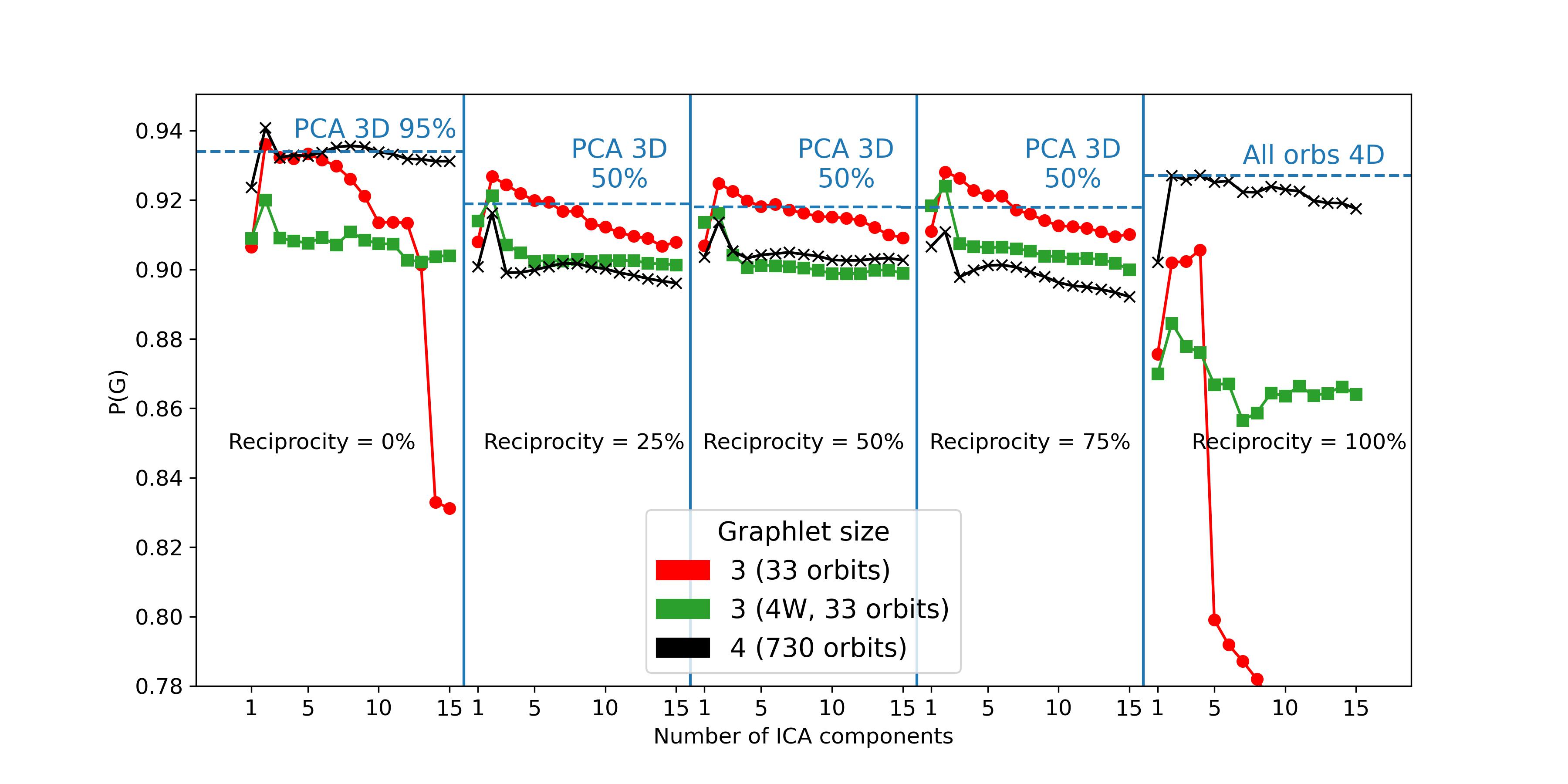}
        \caption{Directed networks - Task 2}
        \label{fig:dir_task2_pbar}
    
        \includegraphics[width=0.77\textwidth]{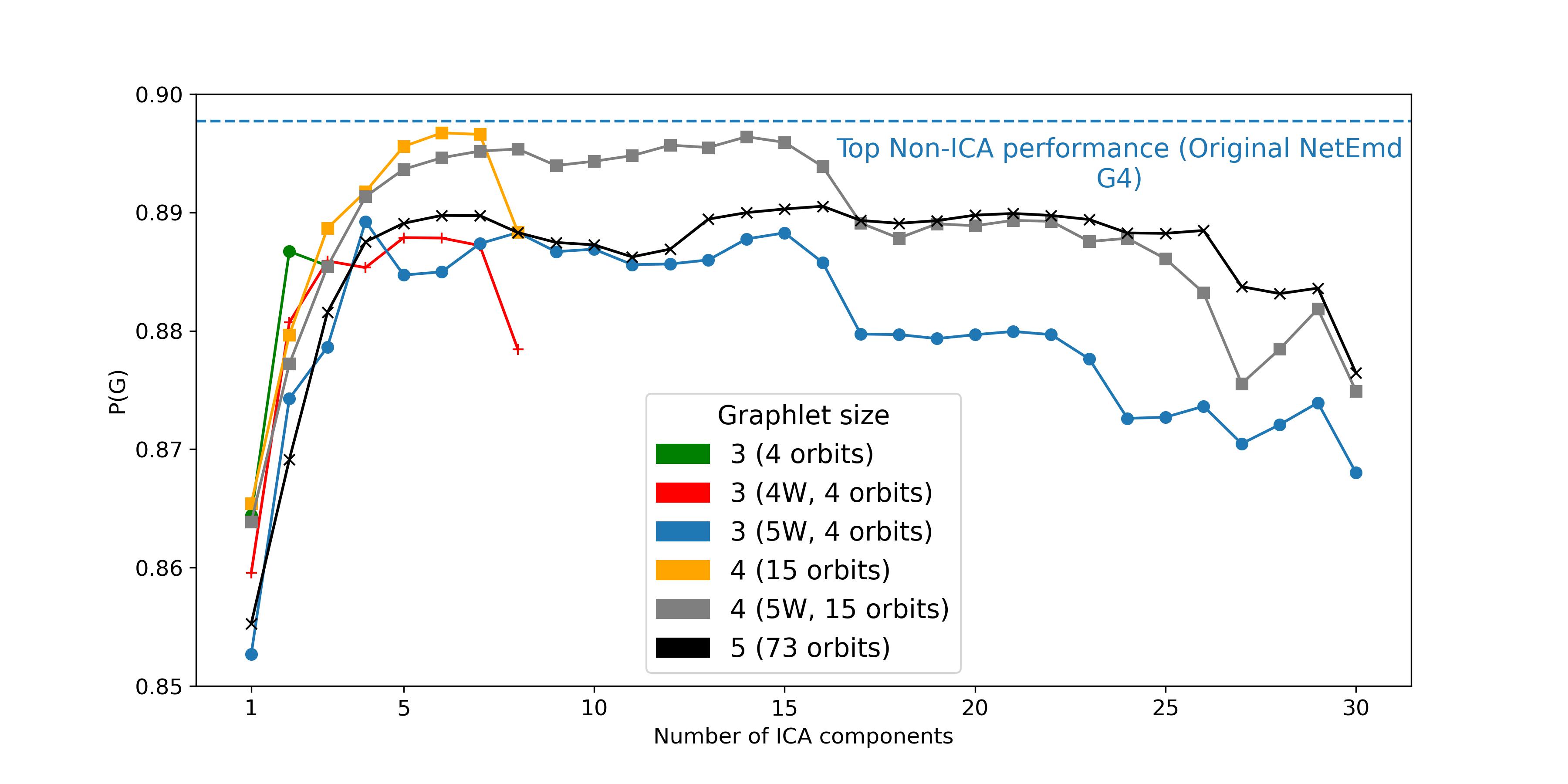}
        \caption{Onnela et al. dataset}
        \label{fig:onnela_pbar}
   
        \includegraphics[width=0.77\textwidth]{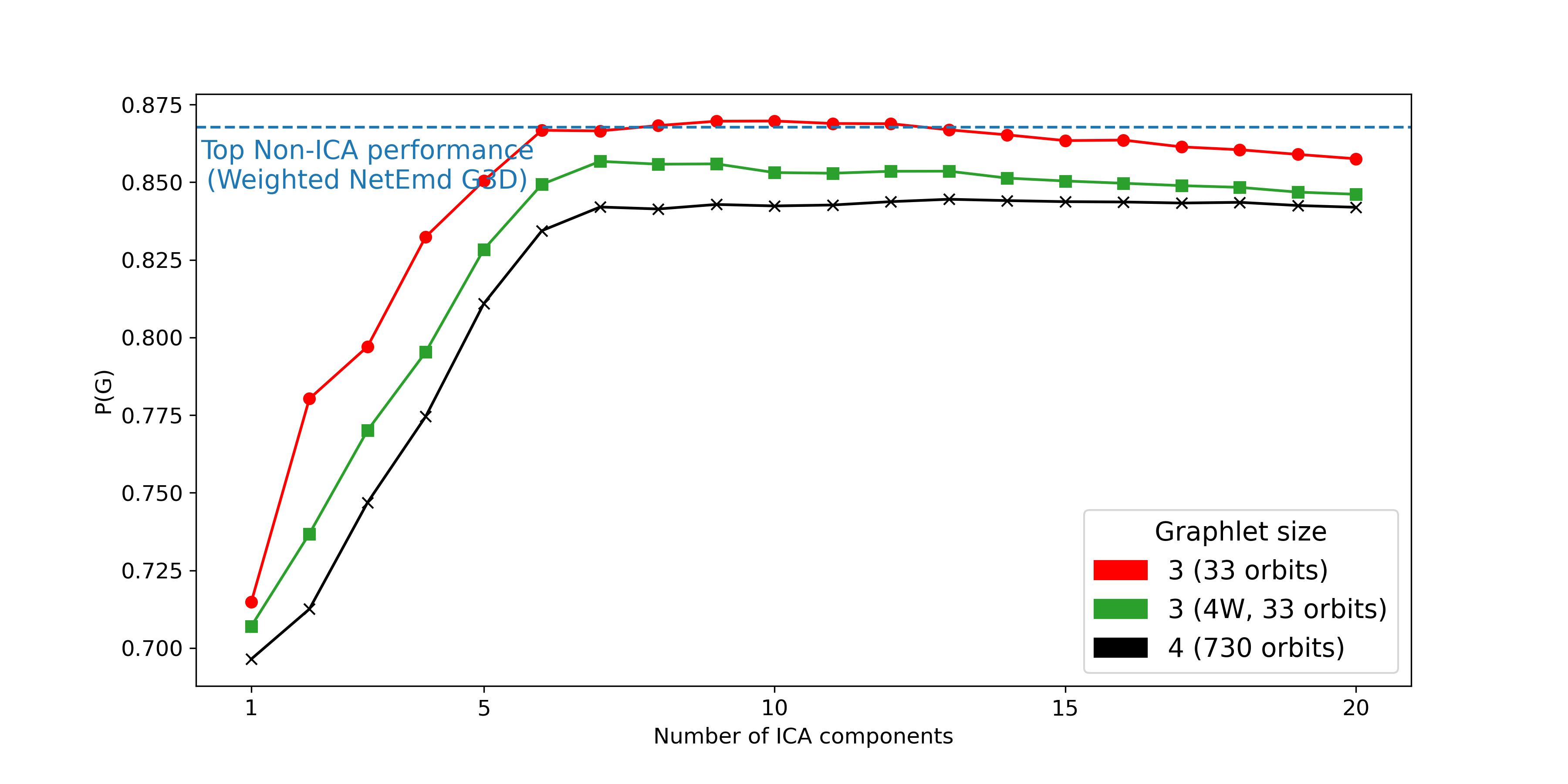}
        \caption{Real world directed networks dataset}
        \label{fig:real_nets_pbar}

\end{figure}

Figures~\ref{fig:undir_task1_pbar} to~\ref{fig:real_nets_pbar} show, in order, performance with varying number of components in undirected networks Task 1, undirected networks Task 2, directed networks Task 1, directed networks Task 2, the Onnela et al. dataset and directed real world networks.

We find that adding more components to ICA does not always translate to a better performance, instead we find that there is a performance maximum obtained with a low number of components that depends on the task. Adding more components after this maximum leads to a decline in performance, either an immediately visible decline (Figures~\ref{fig:undir_task2_pbar},~\ref{fig:dir_task2_pbar} and~\ref{fig:onnela_pbar}) or after plateauing at that maximum value (Figures~\ref{fig:undir_task1_pbar},~\ref{fig:dir_task1_pbar} and ~\ref{fig:real_nets_pbar}). This behaviour indicates that our ICA algorithm is working as expected, as we find a suitable number of components that maximizes the noise reduced in the orbit frequency distributions, so adding more components reduces the difference between $\mathbf{F}_G$ and $\hat{\mathbf{F}}_G$ which negates the positive impact on that noise reduction.

Figures~\ref{fig:dir_task1_pbar} and~\ref{fig:dir_task2_pbar} show a sharp decline in performance when using graphlets of size up to 3 with reciprocity 0\% and 100\%, after 13 and 4 components respectively. This is caused by these extreme levels of reciprocity limiting the orbits that can occur in the networks we compare. With 0\% reciprocity, only graphlets with no reciprocal edges can occur in these networks, so all orbits in graphlets with reciprocal edges have a frequency of 0. As there are only 13 orbits belonging to graphlets with no reciprocal edges, when we try to compute more than 13 independent components the performance degrades as more components are added. With 100\% reciprocity, we find a similar situation but there are only 4 orbits in graphlets of size 2 and 3 for which all edges are reciprocal (orbits 2, 27, 28 and 32 in Figure~\ref{fig:graphlets}).

\section{Conclusion}

We presented two extensions of NetEmd to directed networks, one a direct extension that uses all orbits of graphlets up to size 4 and another that compares networks based only on the orbits that exist in at least one of the networks. We showed that our methodology achieves state of the art performance in large datasets of synthetic networks with heterogeneous network sizes and in datasets of real world networks.

We also proposed to add dimensionality reduction techniques, namely PCA and ICA, as a preprocessing step to the network comparison. The goal of this step is to use dimensionality reduction as a process to attenuate noise within orbit frequencies, that can be introduced by random number generation (in synthetic networks) or data collection/representation (in real world networks). Results show that this preprocessing improves the performance of NetEmd not only when comparing networks with the same number of nodes and edges, but especially in large datasets containing networks of different sizes. 

From an end-user perspective, our extensive testing allows us to recommend guidelines for which version of NetEmd to use in which situation. If the set of networks is homogeneous in number of nodes and average degree, using the largest graphlet size available (size 5 in undirected, size 4 in directed) coupled with ICA (10 to 15 components) leads to the best results. In the cases where there is a large variety in number of nodes or average degree between the networks in the dataset, using smaller graphlets is generally advised. The exception is in directed networks, if the connections within the networks have no reciprocity, then using graphlets of size 4 (with ICA) is more likely to yield a more accurate comparison, due to having a larger set of orbits to with non-zero frequency. On the other hand, if the average reciprocity allows using the whole range of graphlets of size 3 and 4, then it is more likely that using graphlets of size 3 coupled with ICA will lead to a more accurate distinction between networks. In this case, 2 components might be enough if the size of the networks are all within one order of magnitude from each other but 5 to 10 components if there is a wider difference.

The results presented in this contribution indicate that there is value in denoising orbit frequencies when comparing complex networks. However, we only used linear methods to perform the denoising process and graph statistics will not generally depend linearly on underlying parameters. With this in mind, we believe that future work should consider replacing these linear projection methods by others that include a non-linear component, such as non-linear ICA~\cite{hyvarinen2017nonlinear, hyvarinen2019nonlinear} or autoencoder architectures~\cite{kramer1991nonlinear}, such as a variational~\cite{kingma2013auto} or graph autoencoders~\cite{kipf2016variational}.

\section*{Acknowledgments}

MEPS is funded by Engineering and Physical Sciences Research Council Manchester Centre for Doctoral Training in Computer Science (grant number EP/I028099/1). TH is supported by the UKRI through the JUNIPER modelling consortium (grant no. MR/V038613/1) and the Engineering and Physical Sciences COVID-19 scheme (grant number EP/V027468/1), the Royal Society (grant number INF/R2/180067), and the Alan Turing Institute for Data Science and Artificial Intelligence.

\appendix

\section{Discussion on the time complexity of NetEmd}
\label{app:time_complexity}

Calculating the NetEmd measure is a process that can be separated in two phases: acquiring the distributions of the network statistics of interest and comparing the statistics using the EMD. In the case of the algorithm we propose in this work, obtaining the distributions of network statistics involves calculating the graphlet degree matrix and performing PCA or ICA on this matrix.

Obtaining the graphlet degree matrix is the most computationally expensive step of this process, with a time complexity of $O(Nd^{m-1}$), where $N$ is the number of nodes in the network, $d$ the maximum degree of any node in the network and $m$ the size of the graphlets being enumerated. In undirected networks, we use the combinatorial algorithm \emph{ORCA}~\cite{hovcevar2014combinatorial} that relies on an analytical approach to set up a system of linear equations that relates different orbit frequencies. In directed networks, no such approaches are known~\cite{ribeiro2019survey}, so we rely on \emph{G-Tries}~\cite{ribeiro2014g,aparicio2016extending}, a data structure that supports a graphlet enumeration algorithm by representing subgraphs in a prefix tree, the state of the art in enumerating directed graphlets, to our knowledge~\cite{ribeiro2019survey,aparicio2016extending}.

The complexity of PCA is split in two parts, computing the covariance matrix is $O(Np^2)$ and doing the eigenvalue decomposition is $O(p^3)$, where $N$ is the number of nodes in the network and $p$ the number of orbits being considered.  This leads to an overall complexity of $O(Np^2 + p^3)$.

The complexity of ICA is harder to characterize. Firstly, the FastICA algorithm assumes that the data has been centered and whitened~\cite{hyvarinen2000independent}. The implementation we use, from scikit-learn~\cite{scikit-learn}, uses PCA to do the whitening preprocessing, so its complexity is at least as high as PCA. The iterative algorithm to find the weight matrix $\mathbf{W}$ is repeated until the matrix has converged, but there is no prior assurance that the algorithm converges or how many iterations it takes to do so. The scikit-learn implementation defines a maximum number of iterations to stop execution, we set this value to 1000. Each iteration involves calculating $\log\cosh(\mathbf{w}^\intercal\mathbf{f}_i)$ for each component and each node, which the scikit-learn implementation does in $O(Nc^2)$, where $c$ is the number of components to be calculated, and the decorrelation step, which is done in $O(c^2)$. The overall complexity of ICA then becomes $O(Np^2 + p^3 + INc^2)$, where $I$ is the number of iterations.

Wegner et al.~\cite{wegner2017identifying} calculate the complexity of comparing two graphlet distributions using $EMD^*$ to be $O(k(N+N')\log(N+N'))$, where $N$ and $N'$ are the number of nodes of each network and $k$ is the maximum number of function calls to the optimization algorithm used to align the distributions.

\subsection{Runtime analysis}

To illustrate the added time complexity of calculating the linear projections prior to comparing the distributions via the EMD, we show how long each step of the NetEmd process takes for two datasets in Task 1 (one with 1250 nodes and average degree of 10, the other with 10000 nodes and average degree of 80) in directed networks, using graphlet size of 4. Table~\ref{tab:runtime} shows the time taken in minutes to compute the graphlet distributions and the time, to calculate PCA and ICA with different percentages of explained variance or numbers of components and the time to compare the orbit distributions between each pair of networks. These experiments were performed on an AMD Opteron Processor 6380 with 1.4 GHz and 2 MB of cache memory, using Python version 3.5.4, with the \emph{multiprocessing} package and 64 threads. The orbits from graphlets of size 3 and 4 were calculated using \emph{G-Tries}~\cite{ribeiro2014g,aparicio2016extending}, also with 64 threads, whereas the orbits from graphlets of size 2 did not benefit from parallel computation.

\begin{table}[h]
    \centering
    \scriptsize
    \renewcommand{\arraystretch}{1.7}
    \begin{tabular}{>{\centering\arraybackslash}m{50pt}c>{\centering\arraybackslash}m{10pt}>{\centering\arraybackslash}m{8pt}>{\centering\arraybackslash}m{10pt}>{\centering\arraybackslash}m{8pt}>{\centering\arraybackslash}m{5pt}>{\centering\arraybackslash}m{5pt}>{\centering\arraybackslash}m{19pt}}
         \hline & \multirow{2}{29pt}{Orbit distribution} & \multicolumn{3}{c}{PCA} & \multicolumn{3}{c}{ICA} & \multirow{2}{*}{$EMD^*$} \\
         & & 50\% & 90\% & 99\% & 2 & 10 & 20 & \\
         \hline 0\% - (1250,10) & 3.0 & 3.9 & 4.3 & 5.3 & 3.7 & 4.1 & 5.3 & 2.7 \\
         \hline 0\% - (10000,80) & 107 & 22 & 24 & 24 & 22 & 26 & 33 & 21 \\
         \hline 25\% - (1250,10) & 3.0 & 6.0 & 12 & 27 & 5.7 & 5.8 & 6.8 & 3.3 \\
         \hline 25\% - (10000,80) & 121 & 39 & 43 & 54 & 42 & 42 & 48 & 35 \\
         \hline 50\% - (1250,10) & 3.1 & 6.0 & 12 & 27 & 5.9 & 5.9 & 7.3 & 3.3 \\
         \hline 50\% - (10000,80) & 142 & 39 & 43 & 52 & 40 & 41 & 48 & 37 \\
         \hline 75\% - (1250,10) & 3.2 & 6.1 & 12 & 27 & 6.6 & 5.8 & 7.0 & 3.3 \\
         \hline 75\% - (10000,80) & 164 & 38 & 43 & 55 & 40 & 43 & 49 & 36 \\
         \hline 100\% - (1250,10) & 3.3 & 3.6 & 3.6 & 3.6 & 3.5 & 3.8 & 6.2 & 2.6 \\
         \hline 100\% - (10000,80) & 169 & 19 & 20 & 20 & 19 & 23 & 59 & 19 \\
         \hline
    \end{tabular}
    \caption{Time taken in minutes to perform each step of network comparison with directed NetEmd and denoising framework.}
    \label{tab:runtime}
\end{table}

Table~\ref{tab:runtime} highlights the increasing difficulty of obtaining graphlet counts and orbit distributions as the number of nodes and the number of connections in the network increases. In small networks, the time to compute these distributions is less than the time to calculate the linear projections and takes approximately the same time as the comparison via $EMD^*$. However, when increasing the number of nodes by tenfold and the the average degree by eight times, we observe that the computation time of the orbit distributions increases by 30 to 50 times. This growth results in this phase of the algorithm dominating the overall execution time (making up 60\% to 70\% of this time), as the complexity of both the linear projections and $EMD^*$ depends on the number of nodes but not on network density, thus leading to a comparatively smaller increase in their computation time.

This runtime analysis uncovers an unexpected result in the time taken to compute principal components when the amount of explained variance is high. Particularly in small networks, we observe that it takes twice as long to compute the principal components that explain 99\% of variance compared to those that explain 90\% of variance. This sharp increase in execution time is explained by \emph{scikit-learn}'s implementation of PCA, which uses different algorithms depending on the number of components to calculate. It also explains why the execution time of ICA can be lower than PCA, even though ICA uses PCA as a preprocessing step.

Finally, Table~\ref{tab:runtime} demonstrates that calculating up to 10 independent components does not induce a significant increase in the execution time of ICA, when compared with only 2 independent components. However, performing the denoising process with 20 independent components takes on average 33\% longer (1 minute and 30 seconds) in the small networks and 46\% longer (12 minutes) in the big networks than with 10 components; moreover, our results show that there is no visible improvement to the comparison accuracy by doing so.

\section{Results for Task 1 in directed networks}
\label{app:task1_dir}

\afterpage{
\begin{landscape}
\begin{table}
\centering
  \scriptsize
  \renewcommand{\arraystretch}{1.7}
  \begin{tabular}{ccccccc}
    \hline \multirow{2}{*}{Algorithm} & \multirow{2}{*}{Parameter} & \multicolumn{5}{c}{Reciprocity} \\
     &  & 0\% & 25\% & 50\% & 75\% & 100\% \\
    \hline \multirow{2}{*}{All Orbits} & G3D & 0.989 $\pm$ 0.002 & 0.990 $\pm$ 0.002 & 0.989 $\pm$ 0.002 & 0.991 $\pm$ 0.002 & 0.968 $\pm$ 0.003 \\
    \cline{2-7} & G4D & 0.992 $\pm$ 0.002 & 0.993 $\pm$ 0.001 & 0.993 $\pm$ 0.001 & \textit{0.993 $\pm$ 0.001} & \textit{0.989 $\pm$ 0.001} \\
    \hline \multirow{2}{*}{$Weighted\_NetEmd$} & G3D & 0.989 $\pm$ 0.002 & 0.990 $\pm$ 0.002 & 0.989 $\pm$ 0.002 & 0.989 $\pm$ 0.002 & 0.969 $\pm$ 0.003 \\
    \cline{2-7} & G4D & 0.992 $\pm$ 0.002 & 0.993 $\pm$ 0.001 & 0.992 $\pm$ 0.001 & \textit{0.993 $\pm$ 0.001} & \textit{0.989 $\pm$ 0.001} \\
    \hline \multirow{5}{*}{$PCA\_NetEmd$} & 50\% Variance & 0.989 $\pm$ 0.003 & 0.993 $\pm$ 0.002 & 0.991 $\pm$ 0.002 & \textit{0.993 $\pm$ 0.001} & 0.978 $\pm$ 0.004 \\
    \cline{2-7} & 80\% Variance & 0.990 $\pm$ 0.003 & 0.993 $\pm$ 0.001 & 0.990 $\pm$ 0.002 & 0.992 $\pm$ 0.001 & 0.982 $\pm$ 0.003 \\
    \cline{2-7} & 90\% Variance & \textit{0.993 $\pm$ 0.002} & 0.993 $\pm$ 0.001 & 0.992 $\pm$ 0.001 & \textit{0.993 $\pm$ 0.001} & \bf{0.990 $\pm$ 0.002} \\
    \cline{2-7} & 95\% Variance & 0.992 $\pm$ 0.002 & 0.993 $\pm$ 0.001 & 0.992 $\pm$ 0.001 & \textit{0.993 $\pm$ 0.001} & \textit{0.988 $\pm$ 0.001} \\
    \cline{2-7} & 99\% Variance & \textit{0.992 $\pm$ 0.002} & 0.992 $\pm$ 0.001 & 0.992 $\pm$ 0.001 & 0.992 $\pm$ 0.001 & 0.986 $\pm$ 0.002 \\
    \hline \multirow{3}{*}{$ICA\_NetEmd$} & 2 components & 0.990 $\pm$ 0.003 & 0.993 $\pm$ 0.002 & 0.992 $\pm$ 0.002 & \textit{0.993 $\pm$ 0.001} & 0.987 $\pm$ 0.002 \\
    \cline{2-7} & 10 components & \textit{0.992 $\pm$ 0.002} & 0.993 $\pm$ 0.001 & 0.992 $\pm$ 0.002 & \textit{0.993 $\pm$ 0.001} & 0.985 $\pm$ 0.002 \\
    \cline{2-7} & 15 components & \textit{0.993 $\pm$ 0.002} & 0.994 $\pm$ 0.001 & 0.993 $\pm$ 0.001 & \textit{0.993 $\pm$ 0.001} & 0.980 $\pm$ 0.001 \\ 
    \hline
    \hline \multicolumn{2}{c}{TriadEMD} & 0.992 $\pm$ 0.001 & 0.991 $\pm$ 0.001 & 0.990 $\pm$ 0.002 & 0.992 $\pm$ 0.002 & 0.970 $\pm$ 0.003 \\
    \hline \multirow{2}{*}{DGCD} & 13 orbits & 0.986 $\pm$ 0.003 & 0.992 $\pm$ 0.002 & 0.990 $\pm$ 0.002 & 0.988 $\pm$ 0.002 & 0.968 $\pm$ 0.004 \\
    \cline{2-7} & 129 orbits & \bf{0.995 $\pm$ 0.002} & \bf{0.996 $\pm$ 0.001} & \bf{0.995 $\pm$ 0.001} & \bf{0.995 $\pm$ 0.002} & 0.987 $\pm$ 0.002 \\
    \hline \multirow{2}{*}{GDA} & G3D & 0.988 $\pm$ 0.003 & 0.990 $\pm$ 0.002 & 0.988 $\pm$ 0.003 & 0.992 $\pm$ 0.002 & 0.959 $\pm$ 0.005 \\
    \cline{2-7} & G4D & 0.992 $\pm$ 0.003 & 0.992 $\pm$ 0.002 & 0.991 $\pm$ 0.003 & \textit{0.993 $\pm$ 0.002} & 0.977 $\pm$ 0.007 \\
    \hline

  \end{tabular}
  \caption{Results for Task 1 in directed networks, for each level of reciprocity. The metric used for evaluation is the mean (and standard error of the mean) of the 16 values for $\overline{P}$, in each combination of number of nodes, network density and reciprocity. The parameter column indicates: graphlet size used in directed and weighted NetEmd; percentage of variance explained to determine the number of components in $PCA\_NetEmd$ (using orbits in graphlets of size up to 4); the number of components used in $ICA\_NetEmd$, using orbits in graphlets of size up to 4; the number of orbits used by DGCD; graphlet sizes used in GDA. The bolded values are the maximum for each reciprocity level. Italic values are within one standard error of the maximum performance.}
  \label{tab:scores_dir_task1}
\end{table}
\end{landscape}
}

\section{Discussion on orbits from smaller graphlet sizes}
\label{app:smaller_orbits}

Wegner et al.~\cite{wegner2017identifying} have shown that reducing the graphlet size as input to NetEmd can improve the quality of the clusters under certain conditions. A more common reason to use smaller graphlets for comparison is a computational load argument, as we show in Appendix~\ref{app:time_complexity}, increasing the graphlet size leads to more computation time in each step of the NetEmd framework. In the first step, computing orbit frequencies of graphlets of size 4 in directed networks or size 5 in undirected networks can be prohibitive for very large graphs. Another consideration is the time taken to compute the EMD between orbit distributions, particularly in directed networks with graphlets of size 4 requiring 22 times more calls to the $EMD^*$ function than using size 3. Finally, computing the principal or independent components also contain a complexity component dependent on the number of orbits we consider as features to these methods, which also contribute to greater computational load when increasing the graphlet sizes as input to NetEmd. Therefore, it is important to understand in which situations decreasing the graphlet size as input to NetEmd leads to similar or better performance to avoid paying the high computational cost.

We previously presented the performance of $ICA\_NetEmd$ under smaller graphlet sizes, for multiple components used, in Figures~\ref{fig:undir_task1_pbar} to~\ref{fig:real_nets_pbar}. The results for Task 1, both in directed and undirected networks, demonstrate that when comparing networks of similar sizes and densities using bigger graphlet sizes yields a more accurate comparison. The same holds for Task 2 in the undirected case and in the extreme cases of reciprocity (0\% and 100\%) in Task 2 of the directed experiments, where ICA achieves the top performance with size 5 undirected and size 4 directed graphlets, respectively. On the other hand, in the datasets of real world networks and in directed Task 2 with reciprocity of 25\%, 50\% and 75\%, size 5 graphlets in undirected networks and size 4 in directed networks perform worse than size 4 and size 3, respectively. 
For the datasets of real networks in particular, this difference in performance can be explained by noticing that these datasets contain very small networks such that some graphlets have no matches in them. This issue is amplified the larger the graphlet size considered is and it adds a confounding factor to the network comparison, similar to the example of Figure~\ref{fig:weight_by_orbit_example}. Incidentally, this is the only dataset where $Weighted\_NetEmd$ achieves better performance than following the original formulation. 

Figures~\ref{fig:undir_task1_pbar} to~\ref{fig:real_nets_pbar} also show the performance results of using the full set of size $k$ orbits to denoise the distributions of $k-1$ or $k-2$ orbits. In this second way of using ICA, the only time saved compared to using the full set of orbits is the time to compute the EMD between orbit distributions and we find that in no case it leads to the best clustering performance. However, Figure~\ref{fig:undir_task2_pbar} shows a breakpoint at 30 components, where the performance using size 3 and 4 orbits after denoising with size 5 succeeds in performing better than size 5. The reason for this is that, as more components are added, the smaller the reconstruction error $||\mathbf{F}_G - \hat{\mathbf{F}}_G||_2$ is, so the performance with those parameters tends to approximate the performance of the original NetEmd with those graphlets sizes.

The performance of $PCA\_NetEmd$ under smaller graphlet sizes follows a similar set of rules to $ICA\_NetEmd$, with better performances in Task 1 measured when using larger graphlets and in Task 2 and real world networks when using smaller graphlet sizes. Tables~\ref{tab:pca_undir_small} to~\ref{tab:pca_dir_task2_small} detail the results for multiple combinations of graphlet size and percentage of variance explained. Unlike $ICA\_NetEmd$, our experimental setup does not allow us to claim that after a certain value of explained variance performance degrades or stabilizes. Instead, results suggest that the optimal threshold depends on both the dataset and the size of graphlet chosen. Like with $ICA\_NetEmd$, choosing a low value for percentage of explained variance for Task 2 in directed networks yields the best performance for 0, 25, 50 and 75\% reciprocity, but otherwise choosing 90 or 95\% explained variance seems to be a good rule of thumb for best performance with $PCA\_NetEmd$.

\begin{table}[t]
\centering
  \scriptsize
  \renewcommand{\arraystretch}{1.5}
  \begin{tabular}{c>{\centering\arraybackslash}m{60pt}>{\centering\arraybackslash} m{60pt}>{\centering\arraybackslash} m{60pt}}
    \hline \multirow{2}{*}{Parameters} & \multicolumn{2}{c}{Synthetic} & Real \\
    \cline{2-4}  & Task 1 & Task 2 & Onnela et al.\\
    \hline 3 - 50\% Variance & 0.981 $\pm$ 0.004 & 0.912 & 0.872\\
    \hline 3 - 80\% Variance & 0.980 $\pm$ 0.003 & 0.924 & 0.874 \\
    \hline 3 - 90\% Variance & 0.984 $\pm$ 0.002 & 0.929 & 0.881\\
    \hline 3 - 95\% Variance & 0.982 $\pm$ 0.002 & 0.911 & 0.891 \\
    \hline 3 - 99\% Variance & 0.986 $\pm$ 0.002 & 0.928 & 0.889\\
    \hline 4 - 50\% Variance & 0.986 $\pm$ 0.003 & 0.917 & 0.873\\
    \hline 4 - 80\% Variance & 0.987 $\pm$ 0.003 & 0.926 & 0.885 \\
    \hline 4 - 90\% Variance & 0.992 $\pm$ 0.001 & 0.931 & \textbf{0.898}\\
    \hline 4 - 95\% Variance & 0.990 $\pm$ 0.001 & 0.921 & 0.897 \\
    \hline 4 - 99\% Variance & 0.991 $\pm$ 0.001 & 0.931 & 0.897\\
    \hline 5 - 50\% Variance & 0.986 $\pm$ 0.003 & 0.915 & 0.863\\
    \hline 5 - 80\% Variance & 0.985 $\pm$ 0.002 & 0.918 & 0.893 \\
    \hline 5 - 90\% Variance & \textit{0.995 $\pm$ 0.001} & 0.922 & 0.891\\
    \hline 5 - 95\% Variance & 0.993 $\pm$ 0.001 & 0.919 & 0.892 \\
    \hline 5 - 99\% Variance & \textit{0.995 $\pm$ 0.001} & 0.921 & 0.891\\
    \hline Other best & \textbf{0.996 $\pm$ 0.001} (ICA G5 13 components) & \textbf{0.933} (ICA G5 7 components) & \textbf{0.898} (Original NetEmd G4)\\
    \hline
  \end{tabular}
  \caption{Results for $PCA\_NetEmd$ with different combinations of graphlet size and percentage of variance explained in Task 1, Task 2 and the Onnela et al. dataset in the undirected case. The metric used for Task 1 is the mean (and standard error of the mean) of the 16 values for $\overline{P}$ in each combination of number of nodes with network density. The metric used for Task 2 is the sole value of $\overline{P}$ after comparing the 1280 networks. The bolded values are the maximum in each task. Italic values are within one standard error of the maximum performance.}
  \label{tab:pca_undir_small}
\end{table}

\begin{landscape}
\begin{table}
\centering
  \scriptsize
  \renewcommand{\arraystretch}{1.7}
  \begin{tabular}{>{\centering\arraybackslash}m{50pt}>{\centering\arraybackslash}m{60pt}>{\centering\arraybackslash} m{60pt}>{\centering\arraybackslash} m{60pt}>{\centering\arraybackslash} m{60pt}>{\centering\arraybackslash} m{60pt}}
    \hline \multirow{2}{*}{Parameter} & \multicolumn{5}{c}{Reciprocity} \\
    \cline{2-6} & 0\% & 25\% & 50\% & 75\% & 100\% \\
    \hline 3 - 50\% Variance & 0.984 $\pm$ 0.003 & 0.987 $\pm$ 0.003 & 0.985 $\pm$ 0.003 & 0.987 $\pm$ 0.003 & 0.928 $\pm$ 0.004\\
    \hline 3 - 80\% Variance & 0.985 $\pm$ 0.003 & 0.991 $\pm$ 0.002 & 0.991 $\pm$ 0.002 & 0.992 $\pm$ 0.002 & 0.932 $\pm$ 0.004 \\
    \hline 3 - 90\% Variance & 0.989 $\pm$ 0.003 & 0.991 $\pm$ 0.002 & 0.991 $\pm$ 0.002 & 0.990 $\pm$ 0.002 & 0.929 $\pm$ 0.005\\
    \hline 3 - 95\% Variance & 0.992 $\pm$ 0.002 & 0.992 $\pm$ 0.001 & 0.990 $\pm$ 0.002 & 0.992 $\pm$ 0.002 & 0.940 $\pm$ 0.003 \\
    \hline 3 - 99\% Variance & 0.990 $\pm$ 0.002 & 0.991 $\pm$ 0.001 & 0.990 $\pm$ 0.002 & 0.991 $\pm$ 0.002 & 0.938 $\pm$ 0.004\\
    \hline 4 - 50\% Variance & 0.989 $\pm$ 0.003 & 0.993 $\pm$ 0.002 & 0.991 $\pm$ 0.002 & \textit{0.993 $\pm$ 0.001} & 0.978 $\pm$ 0.004\\
    \hline 4 - 80\% Variance & 0.990 $\pm$ 0.003 & 0.993 $\pm$ 0.001 & 0.990 $\pm$ 0.002 & 0.992 $\pm$ 0.001 & 0.982 $\pm$ 0.003 \\
    \hline 4 - 90\% Variance & \textit{0.993 $\pm$ 0.002} & 0.993 $\pm$ 0.001 & 0.992 $\pm$ 0.001 & \textit{0.993 $\pm$ 0.001} & \textbf{0.990 $\pm$ 0.002}\\
    \hline 4 - 95\% Variance & 0.992 $\pm$ 0.002 & 0.993 $\pm$ 0.001 & 0.992 $\pm$ 0.001 & \textit{0.993 $\pm$ 0.001} & \textit{0.988 $\pm$ 0.001} \\
    \hline 4 - 99\% Variance & 0.992 $\pm$ 0.002 & 0.992 $\pm$ 0.001 & 0.992 $\pm$ 0.001 & 0.992 $\pm$ 0.001 & 0.986 $\pm$ 0.002\\
    \hline
    \hline Other best & \textbf{0.995 $\pm$ 0.002} (DGCD129) & \textbf{0.996 $\pm$ 0.001} (DGCD129) & \textbf{0.995 $\pm$ 0.001} (DGCD129) & \textbf{0.995 $\pm$ 0.002} (DGCD129) & \textit{0.989 $\pm$ 0.001} (Weighted G4D)\\
    \hline

  \end{tabular}
  \caption{Results for $PCA\_NetEmd$ with different combinations of graphlet size and percentage of variance explained in Task 1 of directed networks, for each level of reciprocity. The metric used for evaluation is the mean (and standard error of the mean) of the 16 values for $\overline{P}$, in each combination of number of nodes, network density and reciprocity. The bolded values are the maximum for each reciprocity level. Italic values are within one standard error of the maximum performance.}
  \label{tab:pca_dir_task1_small}
\end{table}
\end{landscape}

\begin{table}[h]
\centering
  \scriptsize
  \renewcommand{\arraystretch}{1.7}
  \begin{tabular}{>{\centering\arraybackslash}m{31pt}>{\centering\arraybackslash} m{28pt}>{\centering\arraybackslash} m{28pt}>{\centering\arraybackslash} m{28pt}>{\centering\arraybackslash} m{28pt}>{\centering\arraybackslash} m{28pt}>{\centering\arraybackslash} m{28pt} }
    \hline \multirow{3}{*}{Parameters} & \multicolumn{5}{c}{Synthetic} & \multirow{3}{*}{\shortstack{Real\\Networks}}\\
    \cline{2-6} & \multicolumn{5}{c}{Reciprocity} & \\
     & 0\% & 25\% & 50\% & 75\% & 100\% & \\
    \hline 3 - 50\% Variance & 0.915 & 0.919 & 0.918 & 0.918& 0.857 & 0.825\\
    \hline 3 - 80\% Variance & 0.925 & 0.909 & 0.905 & 0.907 & 0.865 & 0.858 \\
    \hline 3 - 90\% Variance & 0.933 & 0.909 & 0.905 & 0.906 & 0.866 & 0.861\\
    \hline 3 - 95\% Variance & 0.934 & 0.908 & 0.905 & 0.904 & 0.867 & 0.868 \\
    \hline 3 - 99\% Variance & 0.921 & 0.903 & 0.900 & 0.903 & 0.879 & 0.867\\
    \hline 4 - 50\% Variance & 0.924 & 0.897 & 0.900 & 0.894 & 0.908 & 0.821\\
    \hline 4 - 80\% Variance & 0.924 & 0.885 & 0.892 & 0.881 & 0.920 & 0.825\\
    \hline 4 - 90\% Variance & 0.921 & 0.882 & 0.889 & 0.878 & \textbf{0.927} & 0.824\\
    \hline 4 - 95\% Variance & 0.915 & 0.880 & 0.887 & 0.876 & 0.922 & 0.822\\
    \hline 4 - 99\% Variance & 0.914 & 0.878 & 0.885 & 0.875 & 0.924 & 0.818\\
    \hline
    \hline Other best & \textbf{0.941} (ICA G4D 2 comps) & \textbf{0.926} (ICA G3D 2 comps) & \textbf{0.925} (ICA G3D 2 comps) & \textbf{0.928} (ICA G3D 2 comps) & \textbf{0.927} (ICA G4D 4 comps) & \textbf{0.870} (ICA G3D 10 comps) \\
    \hline
  \end{tabular}
  \caption{Results for $PCA\_NetEmd$ with different combinations of graphlet size and percentage of variance explained in Task 2 in directed networks, for each level of reciprocity, and for the real world networks dataset. The metric used for evaluation is the mean (and standard error of the mean) of the 16 values for $\overline{P}$, in each combination of number of nodes, network density and reciprocity. The bolded values are the maximum for each reciprocity level.}
  \label{tab:pca_dir_task2_small}
\end{table}

\section{Different metrics}
\label{app:diff_metrics}

\subsection{Area Under Precision-Recall (AUPR)}

Sarajlic et al.~\cite{sarajlic2016graphlet} evaluate the performance of their graph comparison tool using the Area Under Precision - Recall curve (AUPR), after the framework proposed by Yavero{\u{g}}lu et al.~\cite{yaverouglu2015proper}. The metric is calculated as follows: for each value of a parameter $\epsilon \ge 0$, if the distance between two networks is less than $\epsilon$, then the networks belong to the same cluster. The parameter $\epsilon$ ranges from 0 to 1, with values incrementing by $5e^{-3}$. For each value of $\epsilon$, we compute the number of \emph{true positives}, \emph{false positives}, \emph{true negatives} and \emph{false negatives}. In this context, we define these concepts as:

\begin{itemize}
    \item \emph{True positive (TP)}: the distance between the networks is smaller than $\epsilon$ and they were generated by the same random network model.
    \item \emph{False positive (FP)}: the distance between the networks is smaller than $\epsilon$ but they were generated by different random network models.
    \item \emph{True negative (TN)}: the distance between the networks is greater than $\epsilon$ and they were generated by different random network models.
    \item \emph{False negative (FN)}: the distance between the networks is greater than $\epsilon$ but they were generated by the same random network model.
\end{itemize}

From these quantities, we calculate \emph{precision} as $\frac{TP}{TP + FP}$ and \emph{recall} as $\frac{TP}{TP + FN}$, obtaining 200 tuples of (precision, recall), from which we are able to plot the precision-recall curve. The area under this curve is a metric relevant to our problem as it puts an emphasis on the positive predictive value of the model, disregarding the true negatives that compose the majority of datasets with imbalanced labels such as this one (with 8 random network models, positive labels comprise only $12.5\%$ of the dataset).

\subsubsection{Undirected Results}

Table~\ref{tab:aupr_undir} shows the AUPR for Task 1 and Task 2 in the undirected case and the Onnela et al. dataset.

\begin{table}[t]
\centering
  \scriptsize
  \renewcommand{\arraystretch}{1.7}
  \begin{tabular}{ccccc}
    \hline \multirow{2}{*}{Algorithm} & \multirow{2}{*}{Parameter} & \multicolumn{2}{c}{Synthetic} & Real \\
    \cline{3-5}  &  & Task 1 & Task 2 & Onnela et al.\\
    \hline \multirow{3}{*}{Original NetEmd} & G3 & 0.91 $\pm$ 0.01 & 0.670 & 0.762 \\
    \cline{2-5} & G4 & \textit{0.95 $\pm$ 0.01} & 0.705 & \textbf{0.790} \\
    \cline{2-5} & G5 & \textbf{0.96 $\pm$ 0.01} & 0.712 & 0.761 \\
    \hline \multirow{5}{*}{PCA} & 50\% Variance & 0.93 $\pm$ 0.01 & 0.668 & 0.702\\
    \cline{2-5} & 80\% Variance & 0.92 $\pm$ 0.01 & 0.701 & 0.759 \\
   \cline{2-5} & 90\% Variance & \textit{0.96 $\pm$ 0.01} & 0.717 & 0.756 \\
    \cline{2-5} & 95\% Variance & \textit{0.96 $\pm$ 0.01} & 0.716 & 0.760 \\
    \cline{2-5} & 99\% Variance & \textbf{0.96 $\pm$ 0.01} & 0.718 & 0.760 \\
    \hline \multirow{3}{*}{ICA} & 2 components & 0.93 $\pm$ 0.01 & \textbf{0.742} & 0.717 \\
    \cline{2-5} & 10 components & 0.94 $\pm$ 0.01 & 0.728 & 0.750\\
    \cline{2-5} & 15 components & \textbf{0.96 $\pm$ 0.01} & 0.729 & 0.754 \\
    \hline
    \hline \multirow{2}{*}{GCD} & 11 orbits & 0.89 $\pm$ 0.02 & 0.516 & 0.698\\
    \cline{2-5} & 73 orbits & 0.94 $\pm$ 0.02 & 0.600 & 0.747 \\
    \hline \multirow{3}{*}{GDA} & G3 & 0.92 $\pm$ 0.02 & 0.605 & 0.783 \\
    \cline{2-5} & G4 & 0.80 $\pm$ 0.04 & 0.535 & 0.692 \\
    \cline{2-5} & G5 & 0.74 $\pm$ 0.04 & - & 0.638 \\
    \hline
  \end{tabular}
  \caption{Results for Task 1, Task 2 and the Onnela et al.~\cite{onnela2012taxonomies} datasets in undirected networks. The metric used for Task 1 is the mean (and standard error of the mean) of the 16 values for AUPR in each combination of number of nodes with network density. The metric used for Task 2 and Onnela et al. dataset is the sole value of AUPR after comparing the 1280 and 151 networks, respectively. The parameter column indicates: graphlet size used in original NetEmd; percentage of variance explained to determine the number of components in $PCA\_NetEmd$ (using orbits in graphlets of size up to 5); the number of components used in $ICA\_NetEmd$, using orbits in graphlets of size up to 5; the number of orbits used by GCD; graphlet sizes used in GDA. The bolded values are the maximum for each task. Italic values are within one standard error of the maximum performance. Note that results for GDA with graphlet size 5 took longer than a week to return results for Task 2, at which point we stopped the computation.}
  \label{tab:aupr_undir}
\end{table}

In Task 1, according to AUPR, using PCA or ICA does not lead to a gain in performance, unlike what we observed when using $\overline{P}$. This is not surprising as the improvement we displayed with $\overline{P}$ was within a standard error and both methodologies are able to separate networks between these random models with a very high degree of accuracy. Task 1 serves as a proof of concept that our proposed measure is able to perform the most simple task.

The results for Task 2 are aligned with what we observe when using $\overline{P}$, coupling ICA with NetEmd leads to the best separation between models when different network sizes and densities are present. In the Onnela et al. dataset, the results with AUPR also agree with $\overline{P}$, both performance metrics consider the original NetEmd with size 4 orbits to be one of the top performers with this dataset. The difference between the two metrics is that when measuring with $\overline{P}$, we observe that using PCA with 90\% explained variance leads to similar performance (Table~\ref{tab:pca_undir_small}), but the top performer in this dataset according to AUPR is PCA with 99\% explained variance, achieving an AUPR score of 0.791.

\afterpage{
\begin{landscape}
\begin{table}
\centering
  \scriptsize
  \renewcommand{\arraystretch}{1.7}
  \begin{tabular}{ccccccc}
    \hline \multirow{2}{*}{Algorithm} & \multirow{2}{*}{Parameter} & \multicolumn{5}{c}{Reciprocity} \\
     &  & 0\% & 25\% & 50\% & 75\% & 100\% \\
    \hline \multirow{2}{*}{All Orbits} & G3D & 0.89 $\pm$ 0.01 & 0.90 $\pm$ 0.01 & 0.89 $\pm$ 0.01 & 0.90 $\pm$ 0.01 & 0.914 $\pm$ 0.007 \\
    \cline{2-7} & G4D & \textit{0.961 $\pm$ 0.007} & 0.945 $\pm$ 0.005 & \textbf{0.941 $\pm$ 0.005} & 0.944 $\pm$ 0.005 & \textbf{0.992 $\pm$ 0.002} \\
    \hline \multirow{2}{*}{$Weighted\_NetEmd$} & G3D & 0.87 $\pm$ 0.01 & 0.88 $\pm$ 0.01 & 0.87 $\pm$ 0.02 & 0.89 $\pm$ 0.02 & 0.875 $\pm$ 0.009 \\
    \cline{2-7} & G4D & 0.93 $\pm$ 0.01 & 0.930 $\pm$ 0.007 & 0.925 $\pm$ 0.009 & 0.928 $\pm$ 0.007 & 0.923 $\pm$ 0.009 \\
    \hline \multirow{5}{*}{$PCA\_NetEmd$} & 50\% Variance & 0.94 $\pm$ 0.01 & 0.929 $\pm$ 0.009 & 0.928 $\pm$ 0.008 & 0.927 $\pm$ 0.008 & 0.957 $\pm$ 0.005 \\
    \cline{2-7} & 80\% Variance & 0.94 $\pm$ 0.01 & 0.935 $\pm$ 0.006 & 0.932 $\pm$ 0.008 & 0.932 $\pm$ 0.006 & 0.958 $\pm$ 0.006 \\
    \cline{2-7} & 90\% Variance & \textit{0.958 $\pm$ 0.008} & 0.938 $\pm$ 0.005 & \textit{0.936 $\pm$ 0.005} & 0.936 $\pm$ 0.005 & 0.974 $\pm$ 0.004 \\
    \cline{2-7} & 95\% Variance & \textit{0.958 $\pm$ 0.007} & 0.940 $\pm$ 0.005 & \textit{0.937 $\pm$ 0.005} & 0.939 $\pm$ 0.005 & 0.971 $\pm$ 0.003 \\
    \cline{2-7} & 99\% Variance & \textit{0.958 $\pm$ 0.007} & 0.941 $\pm$ 0.005 & \textit{0.936 $\pm$ 0.005} & 0.940 $\pm$ 0.005 & 0.977 $\pm$ 0.002 \\
    \hline \multirow{3}{*}{$ICA\_NetEmd$} & 2 components & 0.941 $\pm$ 0.007 & 0.938 $\pm$ 0.007 & 0.934 $\pm$ 0.007 & 0.940 $\pm$ 0.007 & 0.982 $\pm$ 0.003 \\
    \cline{2-7} & 10 components & 0.936 $\pm$ 0.009 & 0.941 $\pm$ 0.005 & \textit{0.939 $\pm$ 0.007} & 0.937 $\pm$ 0.006 & 0.936 $\pm$ 0.005 \\
    \cline{2-7} & 15 components & 0.946 $\pm$ 0.007 & 0.938 $\pm$ 0.005 & \textit{0.938 $\pm$ 0.006} & 0.937 $\pm$ 0.005 & 0.92 $\pm$ 0.01 \\ 
    \hline
    \hline \multicolumn{2}{c}{TriadEMD} & 0.938 $\pm$ 0.006 & 0.936 $\pm$ 0.008 & 0.93 $\pm$ 0.01 & 0.94 $\pm$ 0.01 & 0.936 $\pm$ 0.005 \\
    \hline \multirow{2}{*}{DGCD} & 13 orbits & 0.92 $\pm$ 0.02 & 0.93 $\pm$ 0.02 & 0.91 $\pm$ 0.02 & 0.89 $\pm$ 0.02 & 0.87 $\pm$ 0.02 \\
    \cline{2-7} & 129 orbits & \textbf{0.97 $\pm$ 0.01} & \textbf{0.96 $\pm$ 0.01} & \textit{0.94 $\pm$ 0.02} & 0.92 $\pm$ 0.02 & 0.90 $\pm$ 0.02 \\
    \hline \multirow{2}{*}{GDA} & G3D & 0.90 $\pm$ 0.03 & \textbf{0.96 $\pm$ 0.01} & \textit{0.94 $\pm$ 0.01} & \textbf{0.96 $\pm$ 0.01} & 0.82 $\pm$ 0.04 \\
    \cline{2-7} & G4D & 0.88 $\pm$ 0.03 & 0.84 $\pm$ 0.03 & 0.86 $\pm$ 0.03 & 0.84 $\pm$ 0.03 & 0.80 $\pm$ 0.04 \\
    \hline

  \end{tabular}
  \caption{Results for Task 1 in directed networks, for each level of reciprocity. The metric used for evaluation is the mean (and standard error of the mean) of the 16 values for AUPR, in each combination of number of nodes, network density and reciprocity. The parameter column indicates: graphlet size used in directed and weighted NetEmd; percentage of variance explained to determine the number of components in $PCA\_NetEmd$ (using orbits in graphlets of size up to 4); the number of components used in $ICA\_NetEmd$, using orbits in graphlets of size up to 4; the number of orbits used by DGCD; graphlet sizes used in GDA. The bolded values are the maximum for each reciprocity level. Italic values are within one standard error of the maximum performance.}
  
  \label{tab:aupr_dir_task1}
\end{table}
\end{landscape}
}

\subsubsection{Directed Results}

The results using AUPR for Task 1 in directed networks are shown in Table \ref{tab:aupr_dir_task1} and for Task 2 and the dataset of real world directed networks in Table~\ref{tab:aupr_dir_task2}.

\begin{table}[t]
\centering
  \scriptsize
  \renewcommand{\arraystretch}{1.7}
  \begin{tabular}{>{\centering\arraybackslash}m{29pt}>{\centering\arraybackslash}m{43pt}>{\centering\arraybackslash} m{16pt}>{\centering\arraybackslash} m{16pt}>{\centering\arraybackslash} m{16pt}>{\centering\arraybackslash} m{16pt}>{\centering\arraybackslash} m{16pt}>{\centering\arraybackslash} m{28pt} }
    \hline \multirow{3}{*}{Algorithm} & \multirow{3}{*}{Parameter} & \multicolumn{5}{c}{Synthetic} & \multirow{3}{*}{\shortstack{Real\\Networks}}\\
    \cline{3-7} & & \multicolumn{5}{c}{Reciprocity} & \\
     &  & 0\% & 25\% & 50\% & 75\% & 100\% & \\
    \hline  \multirow{2}{*}{\shortstack{All\\Orbits}} & G3D & 0.644 & 0.594 & 0.583 & 0.598 & 0.721 & \textbf{0.883} \\
    \cline{2-8} & G4D & 0.702 & 0.630 & 0.632 & 0.628 & \textbf{0.953} & 0.820 \\
    \hline \multirow{2}{*}{Weighted} & G3D & 0.605 & 0.554 & 0.545 & 0.559 & 0.639 & \textbf{0.883} \\
    \cline{2-8} & G4D & 0.608 & 0.567 & 0.563 & 0.565 & 0.660 & 0.847 \\
    \hline \multirow{5}{*}{PCA} & 50\% Variance & 0.696 & 0.638 & 0.644 & 0.631 & 0.855 & 0.829 \\
    \cline{2-8} & 80\% Variance & 0.700 & 0.626 & 0.634 & 0.621 & 0.845 & 0.824 \\
    \cline{2-8} & 90\% Variance & 0.718 & 0.628 & 0.639 & 0.623 & 0.876 & 0.822 \\
    \cline{2-8} & 95\% Variance & 0.705 & 0.628 & 0.636 & 0.623 & 0.872 & 0.822 \\
    \cline{2-8} & 99\% Variance & 0.703 & 0.628 & 0.637 & 0.623 & 0.892 & 0.819 \\
    \hline \multirow{3}{*}{ICA} & 2 components & \textbf{0.767} & \textbf{0.688} & \textbf{0.691} & \textbf{0.683} & 0.926 & 0.726 \\
    \cline{2-8} & 10 components & 0.725 & 0.669 & 0.679 & 0.664 & 0.820 & 0.837 \\
    \cline{2-8} & 15 components & 0.729 & 0.660 & 0.680 & 0.658 & 0.768 & 0.841 \\
    \hline \multicolumn{2}{c}{TriadEMD} & 0.664 & 0.629 & 0.623 & 0.629 & 0.738 & 0.881 \\
    \hline \multirow{2}{*}{DGCD} & 13 orbits & 0.631 & 0.625 & 0.600 & 0.581 & 0.567 & 0.872 \\
    \cline{2-8} & 129 orbits & 0.653 & 0.643 & 0.620 & 0.593 & 0.558 & 0.860 \\
    \hline \multirow{2}{*}{GDA} & G3D & 0.584 & 0.552 & 0.539 & 0.551 & 0.549 & 0.760 \\
    \cline{2-8} & G4D & - & - & - & - & 0.535 & - \\
    \hline
  \end{tabular}
  \caption{Results for Task 2 in directed networks, for each level of reciprocity, and for the real world networks dataset. The metric used for this task is the sole value of AUPR after comparing the 1280 and 1231 networks, respectively. The parameter column indicates: graphlet size used in directed and weighted NetEmd; percentage of variance explained to determine the number of components in $PCA\_NetEmd$ (using orbits in graphlets of size up to 4); the number of components used in $ICA\_NetEmd$, using orbits in graphlets of size up to 4; the number of orbits used by DGCD; graphlet sizes used in GDA. The bolded values are the maximum for each dataset. Note that results for GDA with graphlet size 4 took longer than a week to return results for Task 2 and in the real world networks dataset, at which point we stopped the computation.}
  \label{tab:aupr_dir_task2}
\end{table}

In Task 1, the main difference we find is that GDA with size 3 orbits shows an improvement over the other algorithms, becoming the top performer with 25\% (tied with DGCD with 129 orbits) and 75\% reciprocity. DGCD with 129 orbits still shows the best performance with 0\% reciprocity, but PCA with 90, 95 and 99\% explained variance is within one standard error, similarly to the results measured with $\overline{P}$. At a 100\% reciprocity, NetEmd with size 4 orbits achieves the best performance, but PCA with 90, 95 and 99\% explained variance and ICA with 2 components obtain similar AUPR scores. As before, we highlight that the high scores achieved by the 6 algorithms in Table~\ref{tab:aupr_dir_task1} serve as a proof of concept that they are able to distinguish the random network models for different levels of reciprocity, even if the AUPR score is not within a standard error of the best score.

In Task 2, similarly to the undirected version, we find an agreement between the AUPR score and $\overline{P}$, with ICA with 2 components highlighted as the highest performing algorithm for 0, 25, 50 and 75\% reciprocity datasets and NetEmd with size 4 orbits for 100\% reciprocity. The same agreement between AUPR and $\overline{P}$ is found on the dataset of real world directed networks, with the top performer being ICA with 6 components with an AUPR of 0.891 but not highlighted in Table~\ref{tab:aupr_dir_task2}.

The conclusions regarding the difference between our proposed NetEmd methods and TriadEMD also hold when using AUPR instead of $\overline{P}$; in the synthetic datasets TriadEMD achieves better AUPR scores than both NetEmd and $Weighted\_NetEmd$ with orbits from size up to 3 graphlets, but worse results than NetEmd with size 4 graphlets. In the real world directed networks dataset, both NetEmd and $Weighted\_NetEmd$ with size 3 orbits obtain better AUPR scores than TriadEMD.

\subsection{Adjusted Rand Index}

The adjusted Rand index~\cite{rand1971objective,hubert1985comparing} is a metric to evaluate the similarity of data clusterings; a corrected for chance version of the Rand index~\cite{rand1971objective} was proposed by Hubert and Arabie~\cite{hubert1985comparing}. Let $\mathcal{G} = \{G_1, G_2, \ldots, G_N\}$ be the set of networks we wish to partition into clusters. For the synthetic datasets, the ground truth partition of this set is known, we have 8 models of random graphs and we evaluate a network comparison method by its ability to reconstruct these  groups. For the Onnela et al.~\cite{onnela2012taxonomies} dataset, the ground truth partition used is the one shown in the supplementary material of~\cite{ali2014alignment}. Let $\mathcal{P}^* = \{P_1^*, P_2^*, \ldots, P_c^*\}$ be the ground truth partitions and $\mathcal{P} = \{P_1, P_2, \ldots, P_c\}$ be the clusters produced by the network comparison algorithms. These clusters are generated by applying a hierarchical clustering algorithm to the matrix of pair-wise distances calculated by the network comparison; in our case we used the implementation by \emph{scikit-learn}~\cite{scikit-learn}, the \emph{AgglomerativeClustering} class with complete linkage.

Given the two set of clusters, we define the following four quantities:

\begin{itemize}
    \item $a$: the number of pairs of elements of $\mathcal{G}$ that are in the same cluster in $\mathcal{P}^*$ and in the same cluster in $\mathcal{P}$.
    \item $b$: the number of pairs of elements of $\mathcal{G}$ that are in different clusters in $\mathcal{P}^*$ and also in different clusters in $\mathcal{P}$.
    \item $c$: the number of pairs of elements of $\mathcal{G}$ that are in the same cluster in $\mathcal{P}^*$ but in different clusters in $\mathcal{P}$.
    \item $d$: the number of pairs of elements of $\mathcal{G}$ that are in different clusters in $\mathcal{P}^*$ but in the same cluster in $\mathcal{P}$.
\end{itemize}

These quantities can be interpreted as the number of true positives, true negatives, false negatives and false positives, respectively, if we consider the attribution of a pair of networks to a cluster as a decision problem. The Rand index is then defined as:

\begin{equation*}
    RI = \frac{a+b}{a+b+c+d} = \frac{a+b}{\binom{N}{2}}.
\end{equation*}
The general form for correcting a index for chance is:

\begin{equation*}
    \frac{RI - \mathbb{E}(RI)}{\text{max}(RI) - \mathbb{E}(RI)}.
\end{equation*}
\label{eq:general_correction}
Hubert and Arabie~\cite{hubert1985comparing} show how to calculate $\mathbb{E}(RI)$, which we omit for brevity, leading to the formula for the adjusted Rand index (assuming a maximum value of 1 for the Rand index when $c$ and $d$ are 0):

\begin{equation*}
    ARI = \frac{\sum_{ij}\binom{n_{ij}}{2} - \frac{\sum_i\binom{a_i}{2}\sum_j\binom{b_j}{2}}{\binom{N}{2}}}{\frac{1}{2} \left[ \sum_i\binom{a_i}{2} + \sum_j\binom{b_j}{2} \right] - \frac{\sum_i\binom{a_i}{2}\sum_j\binom{b_j}{2}}{\binom{N}{2}}},
\end{equation*}
where $n_{ij}$ is the number of networks in common between $P_i^*$ and $P_j$, i.e., $n_{ij} = |P_i^* \cap P_j|$, $a_i = \sum_{j=1}^c n_{ij}$ and $b_j = \sum_{i=1}^c n_{ij}$.

\subsubsection{Undirected Results}

Table~\ref{tab:ari_undir} shows the ARI for Task 1, Task 2 and real world networks in the undirected case.

As with AUPR, we find that using ARI as a performance metric does not change the conclusions drawn from the results with $\overline{P}$. Recalling that the ARI metric measures the quality of hierarchical clusters generated from a distance matrix, for Task 1, the output of original NetEmd with size 5 orbits leads the best score using this metric. More evidence of the quality of the clusters produced by the original NetEmd with size 5 orbits is to notice that increasing the number of components in ICA or the percentage of variance explained in PCA (and therefore approximating the results of original NetEmd) leads to higher ARI score.

For Task 2, the gain measured by $\overline{P}$ and AUPR when using $ICA\_NetEmd$ with 2 components is also reflected in the ARI score. For the Onnela et al. dataset, the best performance we measure not included in Table~\ref{tab:ari_undir} is $ICA\_NetEmd$ with size 4 graphlets and 8 components, with the components calculated with size 5 graphlets, with a score of 0.713.

\begin{table}[ht]
\centering
  \scriptsize
  \renewcommand{\arraystretch}{1.7}
  \begin{tabular}{ccccc}
    \hline \multirow{2}{*}{Algorithm} & \multirow{2}{*}{Parameter} & \multicolumn{2}{c}{Synthetic} & Real \\
    \cline{3-5}  &  & Task 1 & Task 2 & Onnela et al.\\
    \hline \multirow{3}{*}{Original NetEmd} & G3 & 0.63 $\pm$ 0.03 & 0.472 & 0.592 \\
    \cline{2-5} & G4 & 0.69 $\pm$ 0.02 & 0.427 & \textbf{0.675} \\
    \cline{2-5} & G5 & \textbf{0.81 $\pm$ 0.02} & 0.445 & 0.576 \\
    \hline \multirow{5}{*}{PCA} & 50\% Variance & 0.72 $\pm$ 0.03 & 0.439 & 0.558 \\
    \cline{2-5} & 80\% Variance & 0.71 $\pm$ 0.03 & 0.447 & 0.619 \\
   \cline{2-5} & 90\% Variance & 0.78 $\pm$ 0.02 & 0.436 & 0.612 \\
    \cline{2-5} & 95\% Variance & \textit{0.80 $\pm$ 0.02} & 0.414 & 0.610 \\
    \cline{2-5} & 99\% Variance & \textbf{0.81 $\pm$ 0.02} & 0.396 & 0.587 \\
    \hline \multirow{3}{*}{ICA} & 2 components & 0.67 $\pm$ 0.02 & \textbf{0.477} & 0.521 \\
    \cline{2-5} & 10 components & 0.73 $\pm$ 0.03 & 0.452 & 0.593\\
    \cline{2-5} & 15 components & 0.76 $\pm$ 0.03 & 0.469 & 0.602 \\
    \hline
    \hline \multirow{2}{*}{GCD} & 11 orbits & 0.64 $\pm$ 0.07 & 0.309 & 0.435\\
    \cline{2-5} & 73 orbits & 0.77 $\pm$ 0.05 & 0.430 & 0.488 \\
    \hline \multirow{3}{*}{GDA} & G3 & 0.74 $\pm$ 0.05 & 0.361 & 0.456 \\
    \cline{2-5} & G4 & 0.62 $\pm$ 0.07 & 0.332 & 0.197 \\
    \cline{2-5} & G5 & 0.41 $\pm$ 0.05 & -  & 0.207 \\
    \hline
  \end{tabular}
  \caption{Results for Task 1, Task 2 and the Onnela et al.~\cite{onnela2012taxonomies} datasets in undirected networks. The metric used for Task 1 is the mean (and standard error of the mean) of the 16 values for ARI in each combination of number of nodes with network density. The metric used for Task 2 and Onnela et al. dataset is the sole value of ARI after comparing the 1280 and 151 networks, respectively. The parameter column indicates: graphlet size used in original NetEmd; percentage of variance explained to determine the number of components in $PCA\_NetEmd$ (using orbits in graphlets of size up to 5); the number of components used in $ICA\_NetEmd$, using orbits in graphlets of size up to 5; the number of orbits used by GCD; graphlet sizes used in GDA. The bolded values are the maximum for each task. Italic values are within one standard error of the maximum performance. Note that results for GDA with graphlet size 5 took longer than a week to return results for Task 2, at which point we stopped the computation.}
  \label{tab:ari_undir}
\end{table}

\subsubsection{Directed Results}

The results using ARI for Task 1 in directed networks are shown in Table \ref{tab:ari_dir_task1} and for Task 2 and the dataset of real world directed networks in Table~\ref{tab:ari_dir_task2}.

In Task 1, we find that, according to ARI, the performance of all NetEmd variants is significantly below the performance of DGCD with 129 orbits for 0, 25 and 50\% reciprocity and of GDA with size 3 graphlets for 75\% reciprocity. At 100\% reciprocity, NetEmd, $PCA\_NetEmd$ with 80\% variance and GDA, all with size 4 graphlets, share the top performance, but with a smaller difference to other algorithms and parameters. These results are a stark contrast to the ones we observe using AUPR and $\overline{P}$, where scores for this task did not differ significantly between all the algorithms.

In Task 2, unlike in the other metrics, $ICA\_NetEmd$ with 2 components is no longer considered the best performer for 0\% reciprocity, with TriadEMD achieving the best result instead. On the other hand, according to ARI, $ICA\_NetEmd$ with 2 components is the top performer for 100\% reciprocity. For 25, 50 and 75\% reciprocity, $ICA\_NetEmd$ with 2 components performs significantly worse when comparing to the results using AUPR and $\overline{P}$. ARI scores indicate that the best performers in these levels of reciprocity are $PCA\_NetEmd$ with 99\% variance, NetEmd with size 3 graphlets and $PCA\_NetEmd$ with 95\% variance. As we mention previously, using a smaller graphlet size usually leads to better performance when different network sizes and densities are involved. We find partial agreement to this claim when using ARI as the performance metric. At 0\% reciprocity, we find that $ICA\_NetEmd$ with 2 components and size 3 orbits achieves an ARI of 0.509 (higher than the score of $ICA\_NetEmd$ with 2 components and size 4 orbits in Table~\ref{tab:ari_dir_task2}, but lower than the score of TriadEMD); at 75\% reciprocity, $PCA\_NetEmd$ with 90\% variance explained and size 3 orbits achieves an ARI of 0.495 (higher than the score of $PCA\_NetEmd$ with 95\% variance explained and size 4 orbits in Table~\ref{tab:ari_dir_task2}).

For the dataset of real world directed networks, ARI disagrees with AUPR and $\overline{P}$, assigning the best performance to GDA with size 3 orbits. This task also favours using smaller graphlets for the comparison, so the $PCA\_NetEmd$ and $ICA\_NetEmd$ results are at a disadvantage compared to the other methods as we only report results with size 4 in Table~\ref{tab:ari_dir_task2}. For these two algorithms with size 3 graphlets, the best ARI scores on this task are 0.524, when using $PCA\_NetEmd$ with 90\% explained variance, and 0.489, when using $ICA\_NetEmd$ with 19 components.

On the overall, in spite of some agreements between ARI and AUPR or $\overline{P}$, we find ARI to be an unreliable metric to gauge the performance of network distance measures, as it relies on an intermediate step of clustering that is unrelated to the measures themselves. Using a different clustering method or even a different linkage leads to disparate results compared to the ones we present in this section, which adds a confounding factor when measuring the performance of the network comparison itself. An example of this unreliability is the ARI score of 0 given to GDA with size 3 orbits and to TriadEMD at 100\% reciprocity, when the other metrics showed a performance comparable to DGCD.

\afterpage{
\begin{landscape}
\begin{table}
\centering
  \scriptsize
  \renewcommand{\arraystretch}{1.7}
  \begin{tabular}{ccccccc}
    \hline \multirow{2}{*}{Algorithm} & \multirow{2}{*}{Parameter} & \multicolumn{5}{c}{Reciprocity} \\
     &  & 0\% & 25\% & 50\% & 75\% & 100\% \\
    \hline \multirow{2}{*}{All Orbits} & G3D & 0.56 $\pm$ 0.02 & 0.61 $\pm$ 0.03 & 0.62 $\pm$ 0.03 & 0.60 $\pm$ 0.03 & 0.61 $\pm$ 0.02 \\
    \cline{2-7} & G4D & 0.65 $\pm$ 0.03 & 0.70 $\pm$ 0.02 & 0.69 $\pm$ 0.03 & 0.71 $\pm$ 0.02 & \textit{0.64 $\pm$ 0.02} \\
    \hline \multirow{2}{*}{$Weighted\_NetEmd$} & G3D & 0.55 $\pm$ 0.02 & 0.60 $\pm$ 0.03 & 0.60 $\pm$ 0.03 & 0.59 $\pm$ 0.03 & 0.60 $\pm$ 0.03 \\
    \cline{2-7} & G4D & 0.63 $\pm$ 0.03 & 0.68 $\pm$ 0.02 & 0.68 $\pm$ 0.03 & 0.69 $\pm$ 0.02 & \textit{0.63 $\pm$ 0.03} \\
    \hline \multirow{5}{*}{$PCA\_NetEmd$} & 50\% Variance & 0.67 $\pm$ 0.03 & 0.71 $\pm$ 0.03 & 0.72 $\pm$ 0.03 & 0.71 $\pm$ 0.02 & 0.56 $\pm$ 0.01 \\
    \cline{2-7} & 80\% Variance & 0.67 $\pm$ 0.03 & 0.69 $\pm$ 0.03 & 0.67 $\pm$ 0.02 & 0.71 $\pm$ 0.02 & \textit{0.64 $\pm$ 0.02} \\
    \cline{2-7} & 90\% Variance & 0.69 $\pm$ 0.03 & 0.69 $\pm$ 0.02 & 0.69 $\pm$ 0.03 & 0.70 $\pm$ 0.02 & \textit{0.63 $\pm$ 0.02} \\
    \cline{2-7} & 95\% Variance & 0.65 $\pm$ 0.03 & 0.70 $\pm$ 0.02 & 0.68 $\pm$ 0.02 & 0.71 $\pm$ 0.02 & 0.60 $\pm$ 0.01 \\
    \cline{2-7} & 99\% Variance & 0.66 $\pm$ 0.03 & 0.69 $\pm$ 0.03 & 0.68 $\pm$ 0.02 & 0.70 $\pm$ 0.02 & 0.58 $\pm$ 0.02 \\
    \hline \multirow{3}{*}{$ICA\_NetEmd$} & 2 components & 0.63 $\pm$ 0.03 & 0.67 $\pm$ 0.03 & 0.68 $\pm$ 0.03 & 0.70 $\pm$ 0.03 & 0.57 $\pm$ 0.02 \\
    \cline{2-7} & 10 components & 0.64 $\pm$ 0.02 & 0.69 $\pm$ 0.03 & 0.70 $\pm$ 0.03 & 0.68 $\pm$ 0.02 & 0.56 $\pm$ 0.01 \\
    \cline{2-7} & 15 components & 0.64 $\pm$ 0.02 & 0.70 $\pm$ 0.02 & 0.68 $\pm$ 0.02 & 0.70 $\pm$ 0.02 & 0.58 $\pm$ 0.02 \\ 
    \hline
    \hline \multicolumn{2}{c}{TriadEMD} & 0.69 $\pm$ 0.03 & 0.62 $\pm$ 0.03 & 0.65 $\pm$ 0.03 & 0.64 $\pm$ 0.03 & \textbf{0.65 $\pm$ 0.02} \\
    \hline \multirow{2}{*}{DGCD} & 13 orbits & 0.70 $\pm$ 0.03 & \textit{0.79 $\pm$ 0.03} & \textit{0.77 $\pm$ 0.04} & 0.69 $\pm$ 0.05 & 0.59 $\pm$ 0.04 \\
    \cline{2-7} & 129 orbits & \textbf{0.84 $\pm$ 0.03} & \textbf{0.80 $\pm$ 0.03} & \textbf{0.78 $\pm$ 0.04} & 0.74 $\pm$ 0.05 & 0.58 $\pm$ 0.05 \\
    \hline \multirow{2}{*}{GDA} & G3D & 0.66 $\pm$ 0.04 & \textit{0.79 $\pm$ 0.03} & 0.73 $\pm$ 0.03 & \textbf{0.81 $\pm$ 0.03} & 0.56 $\pm$ 0.05 \\
    \cline{2-7} & G4D & 0.66 $\pm$ 0.06 & 0.63 $\pm$ 0.05 & 0.62 $\pm$ 0.07 & 0.61 $\pm$ 0.06 & \textit{0.64 $\pm$ 0.06}\\
    \hline

  \end{tabular}
  \caption{Results for Task 1 in directed networks, for each level of reciprocity. The metric used for evaluation is the mean (and standard error of the mean) of the 16 values for ARI, in each combination of number of nodes, network density and reciprocity. The parameter column indicates: graphlet size used in directed and weighted NetEmd; percentage of variance explained to determine the number of components in $PCA\_NetEmd$ (using orbits in graphlets of size up to 4); the number of components used in $ICA\_NetEmd$, using orbits in graphlets of size up to 4; the number of orbits used by DGCD; graphlet sizes used in GDA. The bolded values are the maximum for each reciprocity level.}
  
  \label{tab:ari_dir_task1}
\end{table}
\end{landscape}
}

\begin{table}[t]
\centering
  \scriptsize
  \renewcommand{\arraystretch}{1.7}
  \begin{tabular}{>{\centering\arraybackslash}m{29pt}>{\centering\arraybackslash}m{43pt}>{\centering\arraybackslash} m{16pt}>{\centering\arraybackslash} m{16pt}>{\centering\arraybackslash} m{16pt}>{\centering\arraybackslash} m{16pt}>{\centering\arraybackslash} m{16pt}>{\centering\arraybackslash} m{28pt}}
    \hline \multirow{3}{*}{Algorithm} & \multirow{3}{*}{Parameter} & \multicolumn{5}{c}{Synthetic} & \multirow{3}{*}{\shortstack{Real\\Networks}}\\
    \cline{3-7} & & \multicolumn{5}{c}{Reciprocity} & \\
     &  & 0\% & 25\% & 50\% & 75\% & 100\% & \\
    \hline  \multirow{2}{*}{\shortstack{All\\Orbits}} & G3D & 0.432 & 0.423 & \textbf{0.464} & 0.387 & 0.188 & 0.167 \\
    \cline{2-8} & G4D & 0.395 & 0.409 & 0.348 & 0.433 & 0.439 & 0.308\\
    \hline \multirow{2}{*}{Weighted} & G3D & 0.393 & 0.361 & 0.343 & 0.319 & 0.311 & 0.449 \\
    \cline{2-8} & G4D &  0.362 & 0.324 & 0.319 & 0.378 & 0.422 & 0.373 \\
    \hline \multirow{5}{*}{PCA} & 50\% Variance & 0.448 & 0.274 & 0.456 & 0.439 & 0.429 & 0.153 \\
    \cline{2-8} & 80\% Variance & 0.425 & 0.371 & 0.338 & 0.396 & 0.424 & 0.124 \\
    \cline{2-8} & 90\% Variance & 0.415 & 0.460 & 0.350 & 0.392 & 0.448 & 0.192 \\
    \cline{2-8} & 95\% Variance & 0.427 & 0.468 & 0.440 & \textbf{0.475} & 0.445 & 0.151 \\
    \cline{2-8} & 99\% Variance & 0.431 & \textbf{0.481} & 0.394 & 0.451 & 0.434 & 0.159 \\
    \hline \multirow{3}{*}{ICA} & 2 components & 0.451 & 0.261 & 0.336 & 0.403 & \textbf{0.484} & 0.118 \\
    \cline{2-8} & 10 components & 0.399 & 0.419 & 0.397 & 0.284 & 0.417 & 0.282 \\
    \cline{2-8} & 15 components & 0.418 & 0.441 & 0.431 & 0.438 & 0.394 & 0.172  \\
    \hline \multicolumn{2}{c}{TriadEMD} & \textbf{0.539} & 0.467 & 0.436 & 0.392 & 0.001 & 0.297 \\
    \hline \multirow{2}{*}{DGCD} & 13 orbits & 0.428 & 0.445 & 0.420 & 0.264 & 0.154 & 0.436 \\
    \cline{2-8} & 129 orbits & 0.432 & 0.402 & 0.421 & 0.386 & 0.373 & 0.203 \\
    \hline \multirow{2}{*}{GDA} & G3D & 0.354 & 0.255 & 0.178 & 0.270 & 0.000 & \textbf{0.542} \\
    \cline{2-8} & G4D & - & - & - & - & 0.455 & - \\
    \hline
  \end{tabular}
  \caption{Results for Task 2 in directed networks, for each level of reciprocity, and for the real world networks dataset. The metric used for this task is the sole value of ARI after comparing the 1280 and 1231 networks, respectively. The parameter column indicates: graphlet size used in directed and weighted NetEmd; percentage of variance explained to determine the number of components in $PCA\_NetEmd$ (using orbits in graphlets of size up to 4); the number of components used in $ICA\_NetEmd$, using orbits in graphlets of size up to 4; the number of orbits used by DGCD; graphlet sizes used in GDA. The bolded values are the maximum for each dataset. Note that results for GDA with graphlet size 4 took longer than a week to return results for Task 2 and in the real world networks dataset, at which point we stopped the computation.}
  \label{tab:ari_dir_task2}
\end{table}

\section{Details of datasets}
\label{app:network_models}

\subsection{Synthetic datasets}

We generate 16 datasets using the combinations of number of nodes $N \in \left\{1250, 2500, 5000, 10000 \right\}$ and average degree $d \in \left\{ 10, 20, 40, 80\right\}$. Each dataset contains 10 realizations of each model for a total of 80 networks.

\begin{itemize}
\item Erd\H{o}s-R{\'e}nyi (ER) model~\cite{erdHos1960evolution}. A random graph with $n$ is generated by picking $m$ unique edges chosen at random from the $\binom{n}{2}$ possible edges. $m$ was chosen to generate networks with the appropriate density.
\item Barabasi-Albert (BA) preferential attachment model~\cite{barabasi1999emergence}. An initial graph is created with $m$ nodes and new nodes are added iteratively to the network, each new node is connected to $m$ existing nodes, picked randomly with probability proportional to their degree. $m$ was chosen to generate networks with the appropriate density.
\item Geometric random graphs~\cite{gilbert1961random}, using a 3-dimensional square ($D = 3$). Nodes are randomly embedded in a $D$-dimensional space and are connected if the Euclidean distance between them is smaller than a threshold $r$. This threshold parameter was determined by grid search to generate networks with the desired number of edges.
\item Geometric gene duplication model~\cite{higham2008fitting}. Starting with an initial network of 5 nodes embedded in a 3-dimensional space, on each iteration a node is chosen randomly to be duplicated. The duplicated node is placed randomly in the 3-dimensional space at an Euclidean distance of at most 2 from the original node. This process is repeated until the desired number of nodes, and nodes at a distance of $r$ or less at connected. This distance $r$ is chosen to generate a network with the appropriate number of edges.
\item Duplication divergence model of Vazquez et al.~\cite{vazquez2003modeling}. The network grows in two stages. In the first stage, a node $v$ is chosen randomly to be duplicated into a node $v'$, that keeps all the edges from $v$. The nodes $v$ and $v'$ are connected with probability 0.05. In the second stage, one of the duplicated edges ($(v,u)$ or $(v',u)$) is chosen randomly and deleted with probability $q$. This process is repeated until the network grows to the desired number of nodes and $q$ is chosen to generate networks with the appropriate density.
\item Duplication divergence model of Ispolatov et al.~\cite{ispolatov2005duplication}. Starting from a network with a single edge, a node is chosen randomly and duplicated, with the new node keeping each of the original's neighbours with probability $p$. This parameter is chosen to generate networks with the desired number of edges.
\item Configuration model, using the Duplication divergence model of Vazquez et al.~\cite{vazquez2003modeling} as the generator for the graphical degree sequence.
\item Watts-Strogatz (WS) model~\cite{watts1998collective}. Nodes are placed in a ring and connected to their $k$ nearest neighbours in both sides of the ring. Each edge is then rewired with probability 0.05 to a new node selected at random. The parameter $q$ is chosen to generate networks with the appropriate average degree.
\end{itemize}

\subsection{Real world networks}

The summary statistics of the Onnela et al.~\cite{onnela2012taxonomies} and the real world directed networks are given in Table~\ref{tab:summary_stats}.

\begin{table}[!ht]
    \centering
    
  \small
    \renewcommand{\arraystretch}{1.5}
  \begin{tabular}{cccc}
  
        \hline \multicolumn{2}{c}{} & Onnela et al. & Real directed \\
        \multicolumn{2}{c}{\# Networks} & 151 & 1283 \\
        \hline \multirow{3}{*}{Number of Nodes} & Min. & 30 & 9\\
        \cline{2-4} & Median & 918 & 160\\
        \cline{2-4} & Max & 11586 & 62586 \\
        \hline \multirow{3}{*}{Number of Edges} & Min. & 62 & 21 \\
        \cline{2-4} & Median & 2436 & 2428\\
        \cline{2-4} & Max & 232794 & 1614977\\
        \hline \multirow{3}{*}{Density} & Min. & $4.26e^{-5}$ & $3.78e^{-5}$ \\
        \cline{2-4} & Median & 0.015 & 0.095\\
        \cline{2-4} & Max & 0.499 & 0.75 \\
        \hline \multirow{3}{*}{Average degree} & Min. & 1.70 & 2.76 \\
        \cline{2-4} & Median & 5.46 & 25.6\\
        \cline{2-4} & Max & 455 & 726 \\
        \hline \multirow{3}{*}{Reciprocity} & Min. & - & 0.0 \\
        \cline{2-4} & Median & - & 0.533 \\
        \cline{2-4} & Max & - & 1.0 \\
        \hline
    \end{tabular}
    \caption{Summary statistics for the Onnela et al. and real world directed networks datasets.}
    \label{tab:summary_stats}
\end{table}

\end{document}